\documentclass[sn-basic]{sn-jnl}

\usepackage{graphicx}
\usepackage{amssymb,amsmath}

\jyear{2021}%

\theoremstyle{thmstyleone}%
\theoremstyle{thmstyletwo}%

\theoremstyle{thmstylethree}%

\newcommand{\dt}{\ensuremath{D_{\Delta{\rm t}}}}
\newcommand{\dd}{\ensuremath{D_\mathrm{d}}}    					
\newcommand{\kmsMpc}{\ensuremath{{\,\rm km}\,{\rm s}^{-1}\,{\rm Mpc}^{-1}}}
\newcommand{\datatd}{\ensuremath{\boldsymbol{\Delta t}}}
\newcommand{\datalens}{\ensuremath{\boldsymbol{d}_{\rm lens}}}
\newcommand{\datakin}{\ensuremath{\boldsymbol{d}_{\rm kin}}}
\newcommand{\dataenv}{\ensuremath{\boldsymbol{d}_{\rm env}}}
\newcommand{\parlens}{\ensuremath{\boldsymbol{\eta}_{\rm lens}}}
\newcommand{\rslope}{\ensuremath{\gamma'}}
\newcommand{\kext}{\ensuremath{\kappa_{\rm ext}}}
\newcommand{\Ddt}{D_{\Delta{\rm t}}}

\newcommand{\xx}{\boldsymbol{\theta}}
\newcommand{\yy}{\boldsymbol{\beta}}
\newcommand{\grad}{\boldsymbol{\nabla}}
\newcommand{\deflectionangle}{\boldsymbol{\alpha}}
\newcommand{\aap}{A\&A}
\newcommand{\mnras}{MNRAS}
\newcommand{\apj}{ApJ}
\newcommand{\prl}{PRL}
\newcommand{\nat}{Nature}
\newcommand{\apjl}{ApJL}
\newcommand{\apjs}{ApJS}
\newcommand{\jcap}{JCAP}

\newcommand{\prd}{Phys.Rev.D}

\newcommand{\aj}{AJ}
\newcommand{\ssr}{SSR}
\newcommand{\pasp}{PASP}

\newcommand{\helens}{HE0435$-$1223}

\newcommand{\be}{\begin{equation}}
\newcommand{\ee}{\end{equation}}

\begin{document}

\title[Time-delay cosmography in the 2020s]{Strong lensing time-delay cosmography in the 2020s}


\author*[1]{\fnm{Tommaso} \sur{Treu}}\email{tt@astro.ucla.edu}

\author[2,3,4]{\fnm{Sherry H.} \sur{Suyu}}\email{suyu@mpa-garching.mpg.de}
\equalcont{These authors contributed equally to this work.}

\author[5]{\fnm{Philip J.} \sur{Marshall}}\email{pjm@slac.stanford.edu}
\equalcont{These authors contributed equally to this work.}

\affil*[1]{\orgdiv{Department of Physics and Astronomy}, \orgname{University of California}, \orgaddress{\city{Los Angeles}, \state{CA} \postcode{90095}, \country{USA}}}

\affil[2]{\orgname{Max Planck Institute for Astrophysics}, \orgaddress{\street{Karl-Schwarzschild-Str.~1}, \city{Garching}, \postcode{85748}, \country{Germany}}}

\affil[3]{\orgname{Technische Universit\"at M\"unchen}, \orgdiv{Physik-Department},  \orgaddress{\street{James-Franck-Str.~1}, \city{Garching}, \postcode{85748}, \country{Germany}}}

\affil[4]{\orgdiv{Institute of Astronomy and Astrophysics}, \orgname{Academia Sinica}, \orgaddress{\street{11F of ASMAB, No.1, Section 4, Roosevelt Road}, \city{Taipei}, \postcode{10617},  \country{Taiwan}}}

\affil[5]{\orgdiv{Kavli Institute for Particle Astrophysics and Cosmology}, \orgname{Stanford University}, \orgaddress{\street{P.O. Box 20450, MS29}, \city{Stanford}, \state{CA} \postcode{94309}, \country{USA}}}


\abstract{Multiply imaged time-variable sources can be used to measure absolute distances as a function of redshifts and thus determine cosmological parameters, chiefly the Hubble Constant H$_0$. In the two decades up to 2020, through a number of observational and conceptual breakthroughs, this so-called time-delay cosmography has reached a precision sufficient to be an important independent voice in the current ``Hubble tension'' debate between early- and late-universe determinations of H$_0$. The 2020s promise to deliver major advances in time-delay cosmography, owing to the large number of lenses to be discovered by new and upcoming surveys and the vastly improved capabilities for follow-up and analysis. In this review -- after a brief summary of the foundations of the method and recent advances -- we outline the opportunities for the decade and the challenges that will need to be overcome in order to meet the goal of the determination of H$_0$ from time-delay cosmography with 1\% precision and accuracy. }

\maketitle


\section{Introduction}
\label{sec:intro}

When a distant variable source (e.g. a supernova or a quasar) is multiply imaged by a foreground mass distribution (e.g. a galaxy or cluster of galaxies), the multiple images appear offset in time to the observer.  The delay(s) between the leading image and trailing one(s) arise from the combination of two effects. The first one is the difference in length of the optical paths. The second is a general relativistic effect, called the \citet{Shapiro1964} delay, owing to the difference in gravitational potential experienced by the photons along the paths. 

Well before the phenomenon was observed, \citet{Ref64} recognized its utility as a way to measure absolute distances and therefore infer cosmological parameters from the expansion history of the universe. In practice, a measurement of the delay(s) between the leading image and the trailing images(s) provides a measurement of the absolute difference in path length, provided the gravitational potential is sufficiently well known. Knowing the redshift, one can then infer the Hubble constant H$_0$ and other cosmological parameters. 

The prospect of a single-step direct measurement of the Hubble constant H$_0$ is particularly appealing in light of the so-called ``Hubble Tension'' -- the 8\% difference between the determination of H$_0$ from multiple local-universe probes \citep[e.g.,][and references therein]{Riess22} and that inferred by early-universe probes under the assumption of a flat $\Lambda$ cold dark matter (CDM) cosmology \citep[e.g.,][and references therein]{Intertwined2022}. If the tension is real and not arising from unknown systematic errors, it implies that new physics is needed beyond the standard flat $\Lambda$CDM model (e.g. an early dark energy phase altering the expansion history of the universe prior to recombination \citep[e.g.,][]{Knox20}).

The technique -- now known as strong lensing time-delay cosmography (or time-delay cosmography, TDC in short) --  has come a long way since the initial proposal by \cite{Ref64}. Hundreds of multiply imaged quasars \citep{Treu18,Lemon22} and the first examples of multiply imaged supernovae have been discovered \citep{Kelly15,Rodney21}. Techniques have been demonstrated to measure time delays with percent accuracy, given sufficient cadence and photometric precision \citep{Courbin18,Millon20b}. Major breakthroughs have been obtained in the modeling of high-resolution space- and ground-based images to constrain the gravitational potential of the main deflector \citep{Suy++10,Chen19,Birrer15,Shajib20}. Non-lensing data, especially stellar kinematics, has been shown to be crucial to break some of the degeneracies inherent to lens modeling \citep{T+K02b,Suyu14,Birrer16,Birrer20}. The role of the environment and of the mass along the line of sight is much better understood than a decade ago \citep{Suy++10,Greene13,Collett13,McCully17}. 

The improvements in data quality and quantity, and analysis techniques, over the past two decades have allowed time-delay cosmography to reach levels of precision of order of 2\% percent in the determination of the Hubble constant H$_0$  by 2020, based on a sample of just 7 lenses \citep{Shajib20,Wong20,Millon20}. 

In the current decade, attention has turned to investigating systematic uncertainties as well as further improving the precision, with the goal of achieving $\sim1\%$ precision {\it and} accuracy.  Achieving this goal will require larger samples ($\gtrsim40$) of lenses with time delays than currently available, as well as high-angular-resolution imaging and spectroscopy to characterize the gravitational potential of the deflector, which is currently considered the main source of residual systematic  uncertainty \citep{Birrer20}. If all goes to plan, this decade will see first light of Euclid, of the Nancy Grace Roman Space Telescope, and of the Vera C.~Rubin Observatory, which should provide orders of magnitude increases in sample size. In parallel, the recent first light of the James Webb Space Telescope and continued development of adaptive optics capabilities from the ground mean that high-angular-resolution imaging and spectroscopic follow-up will become feasible. 

In sum, the 2020 decade is brimming with opportunities for time-delay cosmography, and major breakthroughs are within reach. However, exploiting the wealth of information expected from this influx of data presents non negligible logistical and modeling challenges. Describing the opportunities and challenges of the decade ahead is the main goal of this review, which  builds on a review published in 2016 in this journal \citep[][hereafter TM16]{Treu16b}, to which we refer the reader for more details on the theoretical background, the history of time-delay cosmography,  and developments up until 2015. We also refer the reader to a general review written in 2018 \citep{Suyu2018}, and one that is currently in preparation as a chapter of a book-size conference proceeding of reviews (Birrer et al. 2022, in prep.). 

This review is organized as follows. In Section~\ref{sec:theory}, we give a very brief theoretical background, aimed mostly at defining the notation and clarifying some of the more subtle aspects and recent developments of TDC. In Section~\ref{sec:history}, we review the history of time-delay cosmography, emphasizing chiefly the developments since 2015. In Section~\ref{sec:2020s}, we focus on the decade ahead, identifying the open issues, opportunities, and challenges, and concluding with some forecasts. Section~\ref{sec:summary} summarizes the main points of this review.

\section{Theoretical background}
\label{sec:theory}

In this section we provide a brief summary of the physics of gravitational lenses \citep{SKW06} and how the observable lens time delays depend on both the cosmological world model and our assumptions about the mass distribution of the deflector.   

\subsection{General considerations}
\label{ssec:general}

\begin{figure*}[!t]
\centering\includegraphics[width=0.96\textwidth]{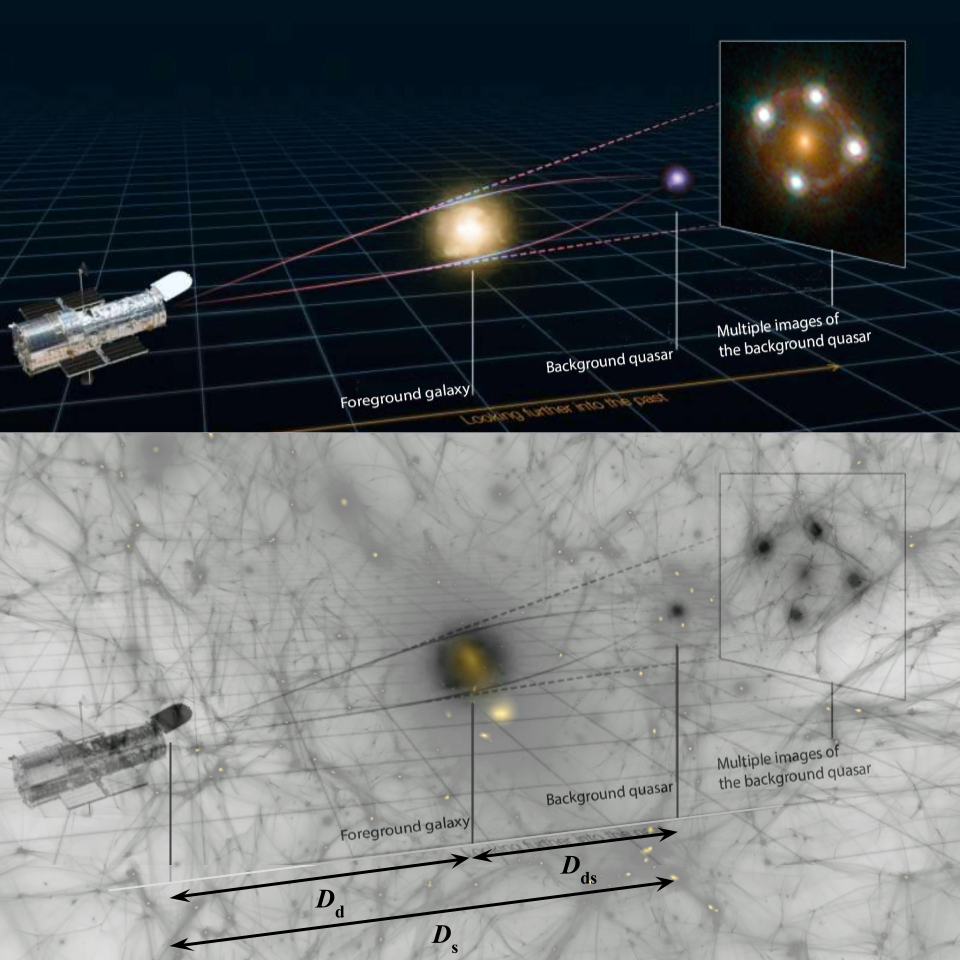}
\caption{Upper panel: schematic lens diagram, illustrating the deflection of light in a galaxy-scale gravitationally lensed quasar system. The deflection is dominated by the massive lens galaxy (graphic credit: Martin Millon). Lower panel: neighboring galaxies of the main deflector, the unseen cosmic web of dark matter that pervades the space along the line of sight to the lens, and the galaxies that inhabit it all contribute weak lensing effects that perturb the light rays and cause systematic error in the time-delay distance inference if they are not accurately accounted for. (The three component distances that go into the time-delay distance are illustrated at the bottom of the panel.) Dark matter simulation and visualization: Tom Abel and Ralf Kaehler.} 
\label{fig:lensdiagram}
\end{figure*}

Gravitational lenses really are lenses, in that the optics of light traveling through the curved space-time around a massive galaxy are identical to those of a glass lens. For example, Fermat's Principle of Least Time for a gravitational lens  \citep{Per90a,Per90b} has that the light travel time through a gravitational lens is
\begin{align}
    \tau(\xx) &= \frac{\Ddt}{c} \cdot \Phi(\xx;\yy), \\
    \text{where\;\;} \Phi(\xx;\yy) &= \frac{1}{2}\left(\xx - \yy\right)^2 - \psi(\xx).
\end{align}
In this equation, $\xx$ and $\yy$ are the apparent (lensed) sky position and
the true (unlensed) sky position of the background source, respectively. 
The  observable position~$\xx$ and the unobservable position~$\yy$ differ by the scaled deflection angle~$\deflectionangle({\xx})$, which is typically $\sim1$~arcsecond in a galaxy-scale strong gravitational lens system.
We recognize the Fermat potential $\Phi(\xx;\yy)$ as the refractive index of the lens. Unlike the refractive index of most glass lenses, $\psi(\xx)$ is, in general, spatially varying: $\psi(\xx)$ is the scaled, projected gravitational potential along the lens sightline. 
While the gravitational potential $\psi(\xx)$ is dominated by the massive deflector, contributions to $\psi(\xx)$ from other structures close to the optical axis have small but significant effects. 
Fig.~\ref{fig:lensdiagram} provides an illustration. 

Multiple images of the background source form at stationary points of the Fermat potential, where $\grad
\tau(\xx) = \grad \Phi(\xx;\yy) = 0$. 
The time delay $\Delta \tau_{\rm AB}$ between brightness fluctuations in image A and image B is observable:
\begin{equation}
    \Delta \tau_{\rm AB} = \frac{\Ddt}{c} \Delta \Phi_{\rm AB}. \label{eq:timedelay}
\end{equation}
Here, $\Delta \Phi_{\rm AB}$ is the Fermat potential difference
between the two image positions, which can be predicted from a model for the mass distribution of the lens, along with the deflection angle $\deflectionangle(\xx)$. Equation~\ref{eq:timedelay} shows the route to inferring cosmological parameters from strong lens time delays: the lens model predicts $\Delta \Phi_{\rm AB}$, and the world model predicts the time-delay distance $\Ddt$, defined in terms of standard angular diameter distances as
\begin{equation}
\Ddt \equiv (1+z_{\rm d}) \frac{D_{\rm d}D_{\rm s}}{D_{\rm ds}},    
\label{eq:Ddt}
\end{equation}
where the subscripts d and s denote the deflector and source, respectively. The predicted time delay can be compared with the observed time delays to infer jointly the lens model and cosmological parameters, as discussed in the next section.

\subsection{From time delays to cosmological parameters}
\label{ssec:TD2COSMO}

\subsubsection{Distance measurements}
\label{ssec:TD2COSMO:dist}

With the lens image, time delays, and lens galaxy's stellar kinematic data, we can constrain the time-delay distance $\dt$ and/or the distance to the lens $\dd$. i.e., we can obtain the marginalized posterior probability distribution $P(\dt)$, $P(\dd)$ and/or $P(\dt,\dd)$ \citep[e.g.,][]{Refsdal64, Paraficz09, Jee19, Birrer16, Birrer19b, Wong20}.  This in turn allows us to measure cosmological parameters, since the distances depend on both the cosmological model parameters and the lens/source redshifts, as we describe below.  

The Friedmann–Lemaître–Robertson–Walker metric that describes a homogeneous and isotropic universe like ours is
\be
\label{eq:FLRWmetric}
{\rm d}s^2 = c^2 {\rm d}t^2 - a^2(t)\left [{\rm d}\chi^2 +f_{K}^2 (\chi)\left({\rm d}\vartheta^2 + \sin^2\vartheta {\rm d} \varphi^2\right) \right ],
\ee
where $c$ is the speed of light, $t$ is the cosmic time, $a(t)$ is the cosmic scale factor (normalized to 1 today at $t_0$, i.e., $a(t_0)=1$), $\chi$ is the comoving radial coordinate, $(\vartheta, \varphi)$ are the angular coordinates on a unit sphere, and $f_{K} (\chi)$ is the comoving angular diameter distance that depends on the spatial curvature $K$ as
\be
\label{eq:fk}
f_{K}(\chi) = \left\{ \begin{array}{ll}
 K^{-1/2} \sin\left(K^{1/2}\chi\right) & \textrm{for $K>0$}\\
 \chi & \textrm{for $K=0$} \\
 (-K)^{-1/2}\sinh\left[(-K)^{1/2}\chi \right] & \textrm{for $K<0$}
  \end{array} \right. .
\ee
The angular diameter distance between two redshifts $z_1$ and $z_2$ is
\be
\label{eq:DA}
D_{\rm A}(z_1,z_2) = \frac{1}{1+z_2}f_{K}[\chi(z_1,z_2)].
\ee

As an example, the expression of $\chi(z_1,z_2)$ in the cosmological model, $\Lambda$CDM, with density parameters $\Omega_{\rm m}$ for matter, $\Omega_{\rm k}$ for spatial curvature, and $\Omega_{\Lambda}$ for dark energy described by the cosmological constant $\Lambda$, is
\be
\label{eq:chi}
\chi(z_1,z_2) = \frac{c}{{\rm H}_0}\int_{z_1}^{z_2} {\rm d}z' \left[\Omega_{\rm m}(1+z')^3 + \Omega_{\rm k} (1+z')^2 + \Omega_{\Lambda} \right]^{-1/2},
\ee
where H$_0$ is the Hubble constant, and $K=-\Omega_{\rm k}{\rm H}_0^2/c^2$ is the spatial curvature.  Consequently, the angular diameter distance $D_{\rm A}$ depends on \{H$_0$, $\Omega_{\rm m}$, $\Omega_{\rm k}$ and $\Omega_{\Lambda}$\} in the $\Lambda$CDM cosmological model.  Since $\dt$ is a combination of angular diameter distances between the observer, lens and source, we can use $P(\dt,\dd)$ to infer cosmological parameters, particularly the Hubble constant.

\subsubsection{The Hubble Constant}
\label{ssec:TD2COSMO:H0}

\begin{figure}[t!]
\centering
\includegraphics[width=0.36\textwidth]{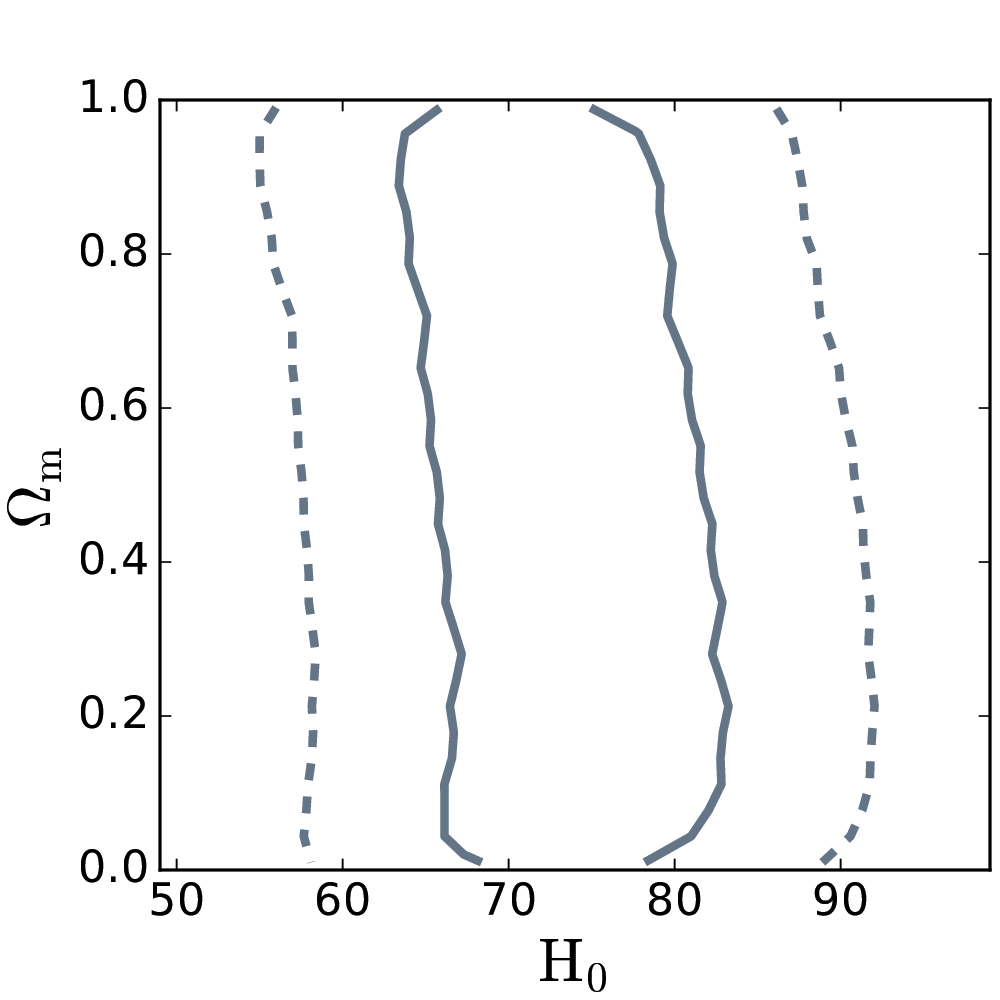} \hspace{0.5cm}
\caption{Constraint on the matter density parameter $\Omega_{\rm m}$ and the Hubble constant H$_0$ in the flat $\Lambda$CDM cosmological model from the $\dt$ measurement to \helens\ \citep{Wong17}.  The contours indicate the 68\% and 95\% credible regions, and the nearly vertical constraint shows that $\dt$ is primarily sensitive to H$_0$. Figure reproduced from \citet{Wong17}.}
\label{fig:he0435H0_DtDdonH0}
\end{figure}

From Equations~(\ref{eq:Ddt}), (\ref{eq:fk}), (\ref{eq:DA}), (\ref{eq:chi}), the time-delay distance $\dt$ is primarily sensitive to H$_0$, in fact inversely proportional.  This is illustrated in 
Fig.~\ref{fig:he0435H0_DtDdonH0} that shows the cosmological constraint on $\Omega_{\rm m}$ and H$_0$ from the $\dt$ measurement to the strongly lensed quasar system \helens\ \citep{Wong17}; the constraint contours are oriented nearly vertically, with most constraints on H$_0$.  The tilt of the contours from vertical depends on the lens and source redshifts, with lower lens redshifts leading to smaller tilts and thus more constraining power on H$_0$.

The distance to the lens, $\dd$, also places constraints on the cosmological parameters, but the dependence on H$_0$ is more correlated with other cosmological parameters.  Therefore, for the same relative uncertainty in distance measurements, $\dt$ provides tighter constraints on H$_0$ than $\dd$.  Nonetheless, the combination of the two distances provides even better constraint on H$_0$ \citep[e.g.,][]{Birrer19b, Wong20, Yildirim20}.  

Since we directly measure distances $\dt$ and $\dd$ from time-delay lensing and kinematics, our inference on H$_0$ depends on the background cosmological model, as evidenced by Equation~\ref{eq:DA}, even though often $\dt$ depends mostly on H$_0$ and only weakly on other cosmological parameters.  Therefore, the H$_0$ measurement from time-delay lensing needs to specify the background cosmological model.  Given the current Hubble tension in flat $\Lambda$CDM model, the H$_0$ measurements from lensing are often quoted for the flat $\Lambda$CDM model for direct comparison.  In order to infer H$_0$ in more general cosmological models with less dependence on the model assumptions, one way is to use the lensing distances to anchor the relative distance scale from supernovae of Type Ia (SNe Ia), as shown by, e.g., \cite{Jee19, Taubenberger19, Arendse20}.

\subsubsection{Beyond the Hubble Constant}
\label{ssec:TD2COSMO:beyondH0}

First and foremost, if the Hubble tension is verified through strong lensing measurements and other cosmological probes, then this has great implications for physics.  We will need new physics beyond the current standard flat $\Lambda$CDM to describe our Universe, which could be, for example, new relativistic species or an early dark energy phase.

In this light, measuring cosmological parameters from lensing distances in these new physical models will be important to shed light on the viable models for our Universe. Even though a single lens system provides mostly constraints on H$_0$, as shown in 
Fig.~\ref{fig:he0435H0_DtDdonH0}, the different tilts of the constraint contours in this plot for different lens/source redshifts reduce cosmological parameter degeneracies. This results in cosmological constraints on parameters in addition to H$_0$ once we have a sample of $\gtrsim$10 lenses at different redshifts.  Furthermore, the cosmological constraints from time-delay lensing are oriented differently in the cosmological parameter space compared to that of other cosmological probes, especially in cosmological models that extend beyond the flat $\Lambda$CDM model.  Therefore, the combination of time-delay lensing with these probes, including SNe Ia \citep{Scolnic18,Brout22} and Cosmic Microwave Background \citep{PlanckCollaboration18}, is powerful in constraining cosmological parameters and models \citep[e.g.,][]{Linder04,Linder11, Suy++10, Jee16, Arendse20, Krishnan21}.

\subsection{The mass-sheet degeneracy}
\label{ssec:MSD}

The above program to infer cosmological models from observed strong lens time delays depends critically on our ability to model accurately the total projected gravitational potential of the lens system. This model is constrained by a number of observables, separate from the time delays. Most important of these observables is the image of the ``Einstein ring,'' which constrains the lens model via the mapping of deflection angles (and thus derivatives of the lensing potential) across the image. This information is complemented by spectroscopy of the deflector galaxy: the kinematics of the stars in the lens galaxy provide independent physical constraints on the lens model. Also important are the observed properties (e.g. position, brightness, color) of the deflector's neighbor galaxies, which inform the modeling of the mass distribution outside the main deflector (whether in the lens plane, or along the line of sight). Even with all possible physical components of the mass distribution included in the model and constrained by observations, some key flexibility remains. Parameter degeneracy, where very similar predictions arise from multiple different parameter combinations, is a generic feature of predictive models. However, lens models contain a particular exact degeneracy that dominates the uncertainty in the time-delay predictions. 

The so-called ``Mass-Sheet Degeneracy'' can be expressed as a transformation of the lens equation involving a rescaling of the source position and deflection angle, plus a corresponding additional deflection equivalent to the lensing effect of a uniform convergence\footnote{Convergence in lensing is used to denote the surface mass density of the lens in units of the so-called critical surface density \cite[e.g.][]{Schneider92}.} field, or ``mass sheet'' \citep{Falco85}:
\be
\label{eq:mst}
\lambda \yy = \xx - \lambda \boldsymbol{\alpha}(\xx) - (1-\lambda)\xx
\ee
The observed image positions are invariant under this transformation,
while the Fermat potentials are re-scaled by the same mass-sheet parameter $\lambda$, such that $\Delta \Phi_{\lambda} = \lambda \Delta\Phi$ (and therefore, for a given cosmology, $\Delta t_{\lambda} = \lambda \Delta t$).

\begin{figure*}[!t]
\centering\includegraphics[width=0.96\textwidth]{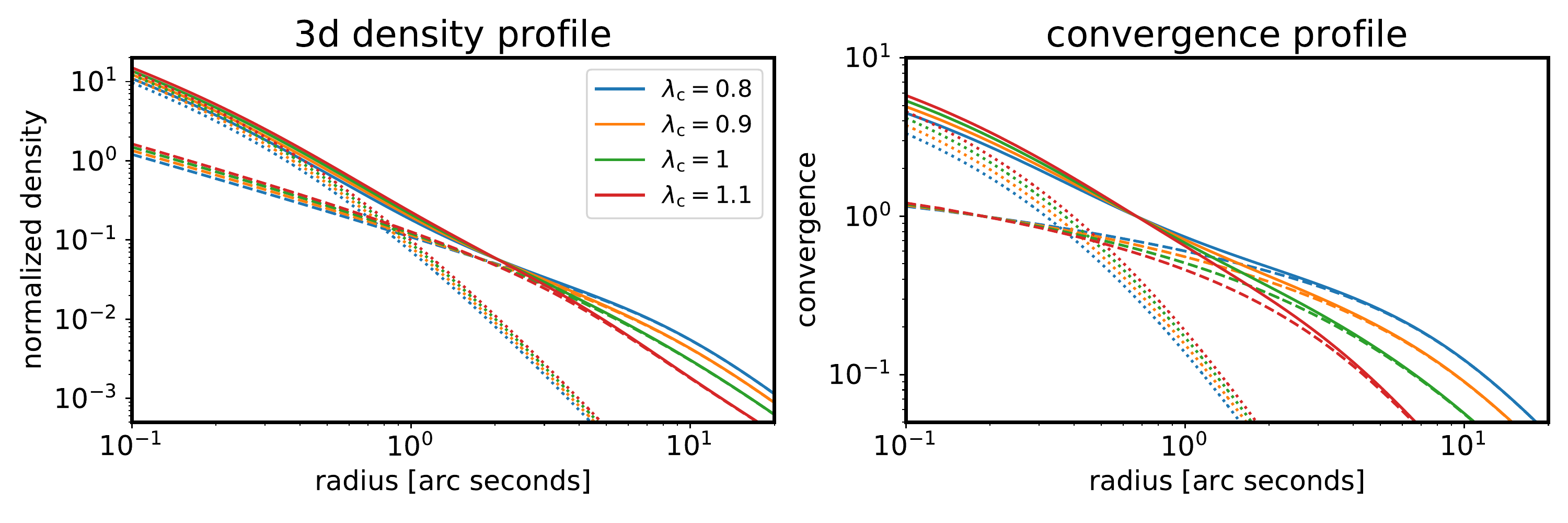}
\caption{Illustration of the mass-sheet degeneracy: a range of deflector mass density profiles yield identical predicted image positions. 
In this figure from \cite{Birrer20}, a composite (stellar, dotted lines, plus dark matter, dashed lines) model lens galaxy has its density profile transformed by an approximate MST (in the notation of the figure $\lambda_{\rm c}$ is the parameter defined as $\lambda$ in this article via Equation~\ref{eq:mst}), leaving the Einstein radius unchanged: imaging data alone cannot distinguish between the four models shown. Non-lensing data, such as stellar kinematics, are required to break the degeneracy. }
\label{fig:msd-H0}
\end{figure*}

The convergence of the lens transforms as follows:
\be
\label{eq:mst-kappa}
\kappa_{\lambda}(\xx) = \lambda \kappa(\xx) + (1-\lambda)
\ee
In practice, lenses are modeled by assuming some functional form for $\kappa(\xx)$, and then inferring the parameters of that function.  The mass-sheet transformation (MST) in Equation~\ref{eq:mst-kappa} shows that there will be sets of model parameter combinations that are related to each other by (approximate) mass-sheet transformations with various values of $\lambda$, which all yield the same (or very similar) observed images but which predict different Fermat potentials/time delays (see examples in Fig.~\ref{fig:msd-H0}).

\begin{figure*}[!t]
\centering\includegraphics[width=0.96\textwidth]{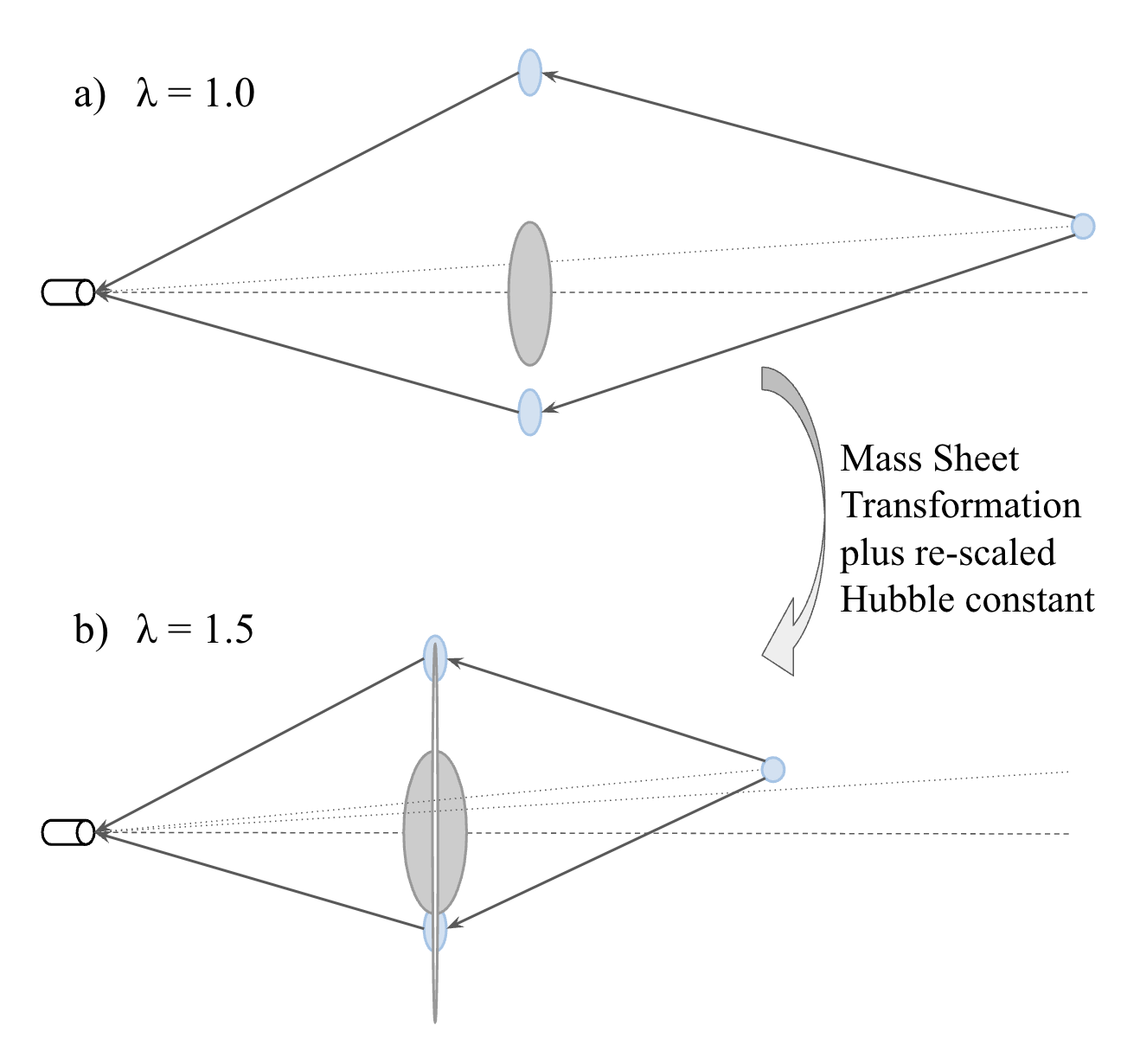}
\caption{Simultaneously applying a mass-sheet transformation to a lens model and re-scaling the Hubble constant by the same transformation parameter $\lambda$ yields a model with both unchanged image positions and time delay. An extreme case of $\lambda = 1.5$ is shown: under this MST, the lens model becomes 50\% more massive but has a negative uniform convergence added to it, and the source position 
has to move 50\% farther from the optical axis (in angle) while the distances need to be scaled downwards by 50\%
in order to keep the image positions fixed and predict the same time delay as in the un-transformed model.} 
\label{fig:mst-rays}
\end{figure*}

Propagating the transformed quantities through the equations that predict the time delay, it can be shown that the inferred time-delay distance and Hubble constant transform as
\begin{align}
\label{eq:Dt_mst}
    D_{\Delta {\rm t},\lambda} &= \lambda^{-1} \Ddt, \\
    H_{0,\lambda} &= \lambda H_0.
\end{align}
Applying an MST with $\lambda > 1$ to a gravitational lens model leaves the image positions unchanged but moves the source angular position $\lambda$ times farther from the optical axis. 
If the model time-delay distance is also scaled down by a factor of $\lambda$ (bringing the observer, lens and source closer together)
this exactly offsets the effect of the MST on the predicted time delay (see Fig.~\ref{fig:mst-rays} for an illustration). 
This model will now fit both the image position data and the time-delay data just as well as the un-transformed model. This is the source of the potential systematic error on H$_0$: the wrong distance can be inferred when jointly fitting observed image positions (or Einstein Ring images) and measured time delays. The MSD can be broken with information about the absolute size or luminosity of the source, or with non-lensing information about the mass distribution in the system.

One way to break the mass-sheet degeneracy is to assume a simply parameterized density profile for the deflector. In this approach, in general, the set of parameter combinations related to each other by the MST is quite small (e.g. because the mass density profile of these models typically have specific shapes such as a power-law radial profile). To put it another way, applying the MST to one of these density profiles will yield a new profile that is not in the assumed model family: it is not possible to approximate the transformed density profile with the assumed functional form. We might consider such an approach to be ``assertive,'' in the sense that it involves asserting that we have very good prior information in favor of that particular density profile functional form. 

At the opposite extreme, we might consider a ``conservative'' approach of including the MSD parameter $\lambda$ in the model explicitly, along with some additional model flexibility, and then marginalizing over it. We consider this approach conservative because the model flexibility is maximally degenerate with distances and therefore H$_0$, equivalent to assuming we know nothing about the mass profiles of galaxies. Prior probability densities for $\lambda$ and the additional model components will be required, perhaps learned from simulations or other datasets, such as for example non time delay lenses \citep{Birrer20}. 

In conclusion, in either the assertive or conservative approach, additional information is needed to constrain the degenerate lens models. The information can be in the form of theoretically or empirically motivated priors or assumptions, or on non-lensing data such as stellar kinematics.

\section{History of time-delay cosmography}
\label{sec:history}

\subsection{Pioneering days: 1964-1999}
\label{ssec:pioneer}

The suggestion by \citet{Ref64} that gravitational lens time delays could be
used to measure a physical distance, and
therefore the Hubble Constant, preceded the discovery of the first multiply imaged quasar \citep{Walsh79} by 15 years. 
Detection of lensed quasars and monitoring of their light curves continued in the eighties and nineties, culminating in the first robust time-delay measurements  \citep{Kun++97,Sch++97}. 
The measurement of time delays was particularly controversial during the nineties as the quality of the early data allowed for multiple estimated values \citep{PRH92a,PRH92b}, owing to the combined effects of gaps in the data, and microlensing noise in the optical light curves. 
This problem was solved definitively at the turn of the
millennium with the beginning of modern monitoring campaigns,
characterized by high cadence, high precision, and long duration, both
at optical and radio wavelengths
\citep{Fas++99,Fas++02,Bur++02,Hjo++02,Eigenbrod05}. 

As robust time delays started to become available, the focus of the controversy shifted to the modeling of the mass distribution of the lens.
Before the year 2000, the only available model constraints were the quasar image positions, time delays, and to a lesser extent the flux ratios (limited by microlensing, variability, and differential extinction). 
Thus, the best one could do was to assume a very simple form for the lens mass distribution, such as a singular isothermal sphere \citep{K+F99}, and to neglect the effects of structure along the line of sight. 
As a result of these necessary but overly simplistic assumptions, the apparent random errors grossly underestimated the total uncertainty, leading to measurements that were apparently significantly inconsistent between groups or with those from other techniques \citep{K+S04}. At the opposite end of the spectrum, pixellated or ``free-form'' models that attempted to reconstruct the mass distribution or gravitational potential on a grid were underconstrained by the data and therefore their results and estimated uncertainties depended strongly on the regularization scheme and prior \citep{WS2000}.

Since recognizing these problems, the community has been pursuing high-quality data for each lens system, as discussed in the next Section.

\subsection{Modern times: 2000-2020}
\label{ssec:modern}

We focus here on cosmography with lensed quasars in the past two decades.  New types of strongly lensed transients, such as lensed supernovae \citep[e.g.,][]{Kelly2015, Goobar17,Rodney21} and lensed gamma-ray bursts \citep[e.g.,][]{Paynter21}, are starting to be detected in recent years and will be discussed in Section \ref{sec:2020s}.  

To infer the distances $\dt$ and $\dd$, we need to model the lens mass distribution, particularly to obtain the Fermat potential at the quasar image positions where we measure the time delays. In the past decade, new imagers and spectrographs on space-based and ground-based telescopes have enabled the acquisition of high-SNR and high-resolution data of lens systems, particularly the time delays $\datatd$ (Section \ref{ssec:modern:delays}), imaging data $\datalens$, kinematic data $\datakin$ of the foreground deflector (Section \ref{ssec:modern:modeling}), and environment data $\dataenv$ of the lens (Section~\ref{ssec:modern:los}).  We first describe how lenses are found in Section~\ref{ssec:modern:sample} before diving into the assembly of the necessary ingredients and analysis in Sections~\ref{ssec:modern:delays}-\ref{ssec:modern:los}, and we present the latest cosmographic results from the past two decades of efforts in Section~\ref{ssec:modern:cosmo}.

\subsubsection{Finding new lenses}
\label{ssec:modern:sample}

Strongly lensed quasars are inherently rare (and the other lensed transients with shorter durations of the transients are even rarer), with about 1 in $\sim$10$^4$ massive galaxies acting as a lens.  After the first discoveries of lensed quasars \citep[e.g.,][]{Walsh79}, the Cosmic Lens All-Sky Survey \citep[CLASS;][]{Mye++03, Bro++03} systematically searched for lensed quasars by looking for flat spectrum radio sources with multiple components, which provided the first large sample of $\sim$20 lensed quasars.  

The next large sample came from the systematic search through the SDSS Quasar Lens Search \citep[SQLS;][]{Oguri06, Inada12, More16}, which performed a systematic search through the SDSS quasar sample.  In recent years, wide-field imaging surveys that go deeper with better image quality (e.g. PanSTARRS, DES, ATLAS and HSC surveys, and the space telescope Gaia) have been a treasure trove for lensed quasar searches with now several hundreds of lensed quasars discovered \citep[e.g.,][]{Agnello15, Lemon17, Ostrovski17, krone-martins2018, Agnello18b, Rusu19, Chan20,Lemon22}.

From these new lensed quasars, we can form a cosmological sample by acquiring the necessary ingredients: redshifts, time delays, high-resolution imaging, lens stellar kinematics for mass modeling, and wide-field imaging/spectroscopy data for characterising the environment.  The lens and source redshifts which could either come from the spectroscopic observations used to confirm the nature of the lens candidates or from further spectroscopic observations.  We describe the other ingredients and analysis in the following subsections.

\subsubsection{Measuring time delays}
\label{ssec:modern:delays}

Obtaining time-delay measurements of lensed quasars requires dedicated monitoring. The basic idea is to detect variations in the brightness of the quasar images in a lens system and use these variations to determine the time delay between the multiple images, given that the intrinsic brightness variations of the quasar manifest in each of the multiple images. While simple in theory, practical implementation requires high cadence to detect the often small variations in the quasar brightness. Furthermore, the brightness of the quasars could also change due to microlensing by the stars in the foreground lens galaxy as these stars move past the sightline of the quasar images.  Measuring the delays therefore requires disentangling the intrinsic quasar variability and the extrinsic microlensing variability.

Monitoring has been carried out in radio wavelengths \citep[e.g.,][]{Fas++02,Rumbaugh15} and optical wavelengths \citep[e.g.,][]{Koc++06a, Vui++08, Paraficz09a, Courbin11}.  Radio monitoring has the advantage that microlensing is minimal at these wavelengths given the large source size relative to the Einstein radii of the lensing stars, so the changes in the light curves of each quasar image mostly originate from intrinsic quasar variability.  However, variability of quasars are typically small in the radio wavelengths, making it more difficult to detect features in the light curves to measure delays \citep{Rumbaugh15}.  Despite this, it was with radio monitoring that \citep{Fas++02} measured all three independent time delays of the quad lens system CLASS\,B1608+656 with uncertainties of a few percent, making this the first system that was used for accurate cosmological distance measurements \citep{Suy++10, Jee19}. 

Most of the monitoring campaigns in the past decade have used optical facilities, notably by the COSmological Monitoring Of GRAvItational Lenses \citep[COSMOGRAIL;][]{Courbin11} using a network of 1-2m class telescopes across the globe and targetting the brighter lensed quasar systems.  In the optical, the microlensing effects on the light curves are often prominent, and disentangling intrinsic and extrinsic variability has traditionally required decade-long light curves to obtain accurate time-delay measurements \citep{Tewes13}.  While COSMOGRAIL has monitored lenses in the last two decades, it is difficult to scale up on the number of lenses for cosmography in this fashion.  The breakthrough in the past decade achieved by COSMOGRAIL came from the following ingredients, which allowed measurement of delays within 1-2 years of monitoring: i) high-cadence (daily) observations, ii) high signal-to-noise (SNR) ratio ($\gtrsim1000$) of the lensed quasars images with milli-mag uncertainties on the photometries, and iii) new curve-shifting and delay measurement techniques that incorporate the uncertainties due to microlensing.

\begin{figure}[t!]
\centering
\includegraphics[width=0.9\textwidth]{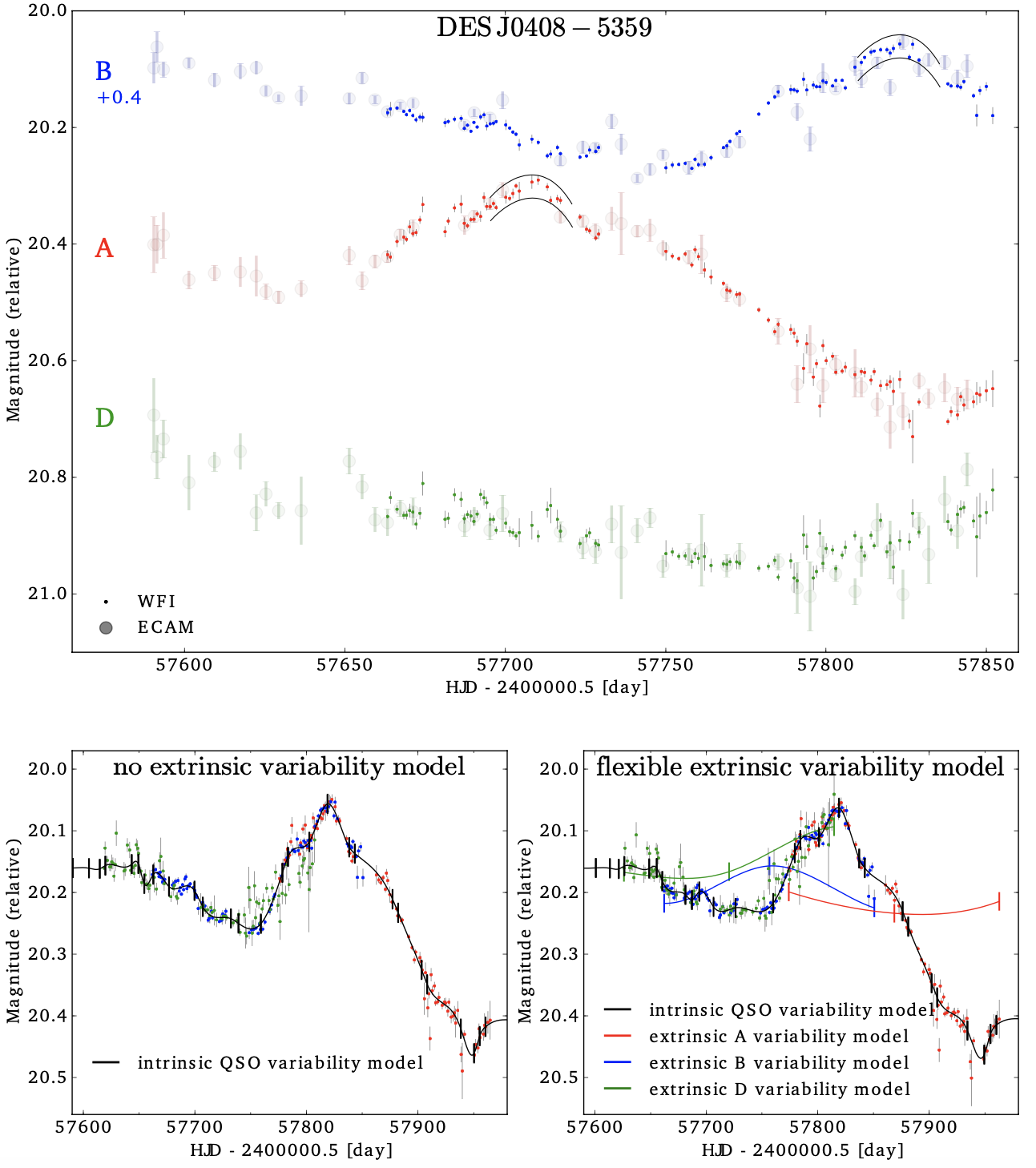}
\caption{Top panel: light curves of the three brightest quasar images A, B and D of the strong lens system DES\,J0408$-$5359 over merely one monitoring season.  The solid circular data points correspond to the high-cadence and high-SNR observations from the WFI instrument on the ESO/MPG 2.2m telescope, whereas the large faint data points correspond to the measurement from the 1.2m Euler telescope.  The black lines show high-frequency variability in image A and B that allow the delays to be measured with only one season of monitoring.  Bottom left (right) panel: time-shifted light curves of each of the three quasar images without (with) corrections for microlensing distortions.  By modeling the extrinsic variability of each light curves due to microlensing, the bottom right panel shows a good match in the intrinsic quasar light curves of the 3 images.  \citep[reproduced from][]{Courbin18} }
\label{fig:des0408lightcurve}
\end{figure}

Figure~\ref{fig:des0408lightcurve} is an example of high-cadence and high-SNR monitoring fulfilling i) and ii) of the lensed quasar system DES0408$-$5359, showing the light curves of the three brightest quasar images (A, B and D).  The small data points are the high-cadence and high-SNR ones from the WFI instrument at the ESO/MPG2.2m telescope, whereas the larger data points are more sparse and noisier from the smaller 1.2m Euler telescope \citep{Courbin18}.  The high-cadence and SNR data points show high frequency variations (e.g., the structure between the black solid lines) that can only come from quasar intrinsic variability and not microlensing, effectively allowing the time delays to be measured within a single year of monitoring.

Various techniques have been developed in the past decade to infer time delays, including spline fitting in the Python Curve Shifting package \citep[PyCS;][]{Tewes13b}, Gaussian processes \citep{Hojjati13}, smoothing and cross-correlation \citep[e.g.,][]{AghamousaShafieloo15}, and Bayesian curve fitting with a latent continuous-time Ornstein-Uhlenbeck process \citep{Tak+16}.  To crash test these approaches for time-delay measurement, a time-delay challenge has been set up where an ``evil'' team simulated mock light curves with the input delays hidden, and subsequently multiple ``good'' teams with various time-delay methods tried to measure the delays from these light curves as part of a blind test of their methods' performances \citep{Dobler15, Liao15}.  The results of the challenge demonstrate that some of the methods reach the target precision and accuracy in the delay measurements for cosmography, particularly the PyCS approach, which has been the basis of the COSMOGRAIL time-delay measurements.

A summary of the time-delay measurements of 37 galaxy-scale lensed quasar systems is presented in Table 3 of \cite{Millon20a}, where the majority of the time delays have been measured by the COSMOGRAIL collaboration through 15 years of monitoring.  In addition to these, the time delays of 6 lensed quasars with the high-cadence and SNR monitoring are presented by \cite{Millon20b}.

\subsubsection{Modeling the mass distribution of the main deflector and its neighbours}
\label{ssec:modern:modeling}

We describe the modeling process to predict the lensing and kinematic observables in order to compute the lensing likelihood $P(\datalens \vert \parlens, \dt, \dd)$ and kinematic likelihood $P(\datakin \vert \parlens, \dt, \dd)$, where $\parlens$ are the lens mass model parameters.  

\begin{figure}[t!]
\centering
\includegraphics[width=0.9\textwidth]{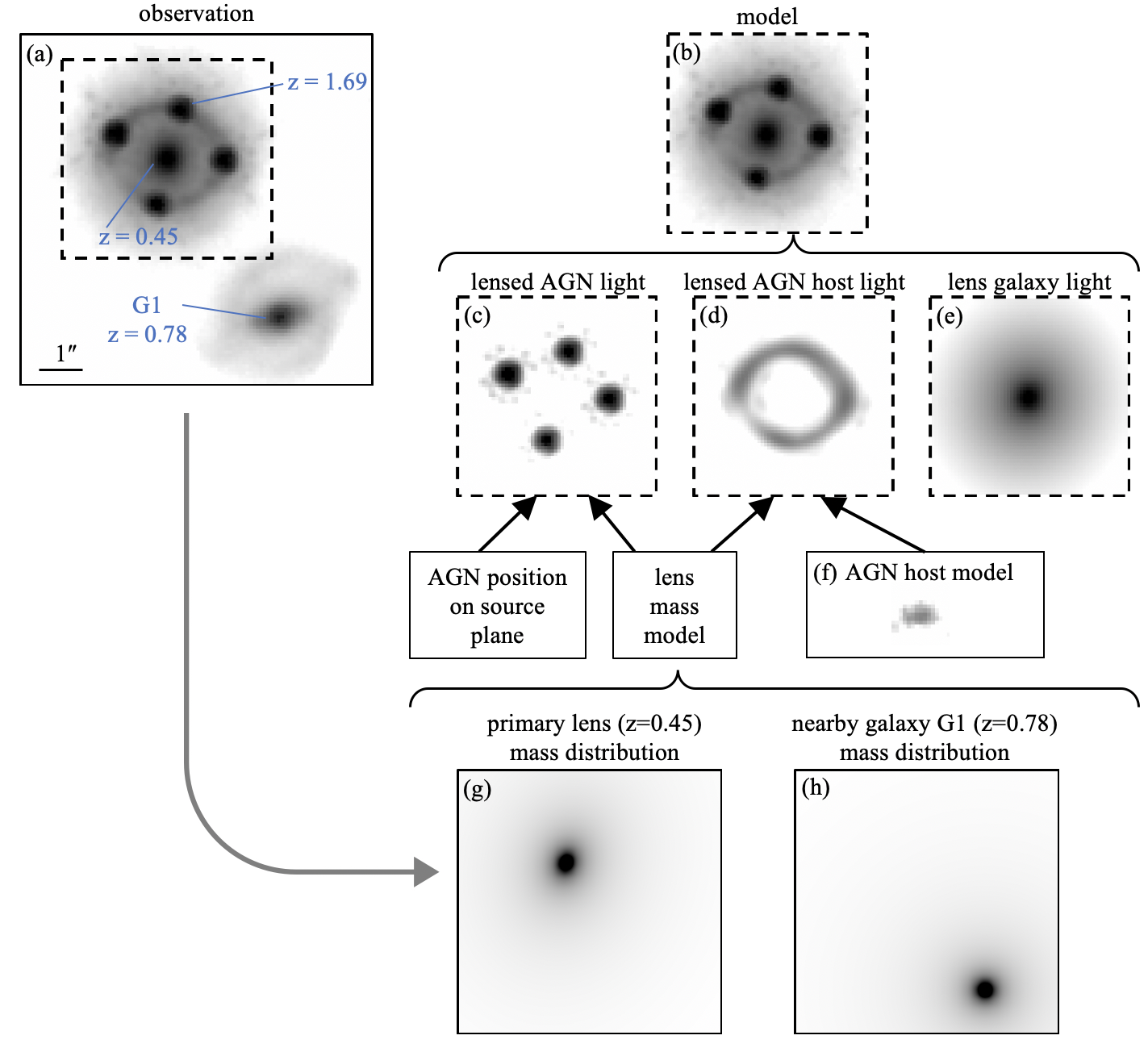}
\caption{Panel (a): high-resolution \textit{HST} image of HE0435. Panel (b): model of HE0435, composed of the three components of the light distribution in panels (c) the lensed AGN light, (d) the lensed AGN host galaxy light, and (e) the primary lens galaxy light. Panel (f): the model AGN host galaxy light distribution.  Producing the light distributions in panels (c) and (d) requires a model of the lens mass distribution, the AGN source position, and the AGN host galaxy light distribution.  Panels (g) and (h): convergence map for the primary lens and the nearby galaxy respectively.  Figure based on mass models from \citet{Wong17}.} 
\label{fig:model_eg}
\end{figure}

Let us consider a high-resolution image of the lens system as the lensing data $\datalens$, with an example shown in Figure \ref{fig:model_eg} panel (a).  Studies in the past decades have shown that the thousands of intensity pixels especially of the lensed quasar host galaxy provide stringent constraints on the lens mass distribution \citep[][]{Kochanek2001,Dye05, Suyu09, Suyu14, Wong17, Chen19}.  In order to use this data set, we need to model the image (panel (b)) with various components of light in the system, which are the lensed AGN (panel (c)), the lensed AGN host galaxy (panel (d)), and the primary deflector (panel (e)).  In order to produce panels (c)-(e), we need to have a good model of the point spread function (PSF) of the telescope, especially for panel (c) with the bright AGN (point sources).  We further require a lens mass model to ensure that a modeled AGN position on the source plane produces the multiple lensed AGN positions on the image plane that match the observations.  The same lens mass model, together with a model of the surface brightness distribution of the AGN host galaxy (panel (f)), is also used to predict the lensed AGN host light shown in (d).  The lens mass model accounts for all deflectors that have significant strong lensing effects on (c) and (d)\footnote{Quantitatively, the flexion shift criterion introduced by \cite{McCully17} can be used to determine which neighboring galaxies near the primary deflector need to be included directly in the strong lens model}.  In this specific example, we show that the primary deflector and the nearest galaxy G1 have their mass distributions included in the strong lens model (panels (g) and (h)).  The modeling procedure therefore requires a high number of parameters to describe the PSF, the deflectors' mass distributions, the primary deflector's light distribution, and the AGN host galaxy light distribution (up to hundreds of parameters, although some can be linear parameters and be optimized/marginalized analytically).

The PSF can be built from stars in the wider field of view (FOV) of the lensing image.  However, due to color differences between the stars and the AGN, and the often limited number of stars in the FOV, the PSF model built from stars rarely match the AGN light down to the noise level.  Therefore, modeling methods have been developed in recent years to simultaneously reconstruct the PSF together with the lens mass model by making use of the multiple AGN images \citep[e.g.,][]{Chen16, Wong17,Birrer17}. 

\begin{figure*}[!t]
\centering\includegraphics[width=0.96\textwidth]{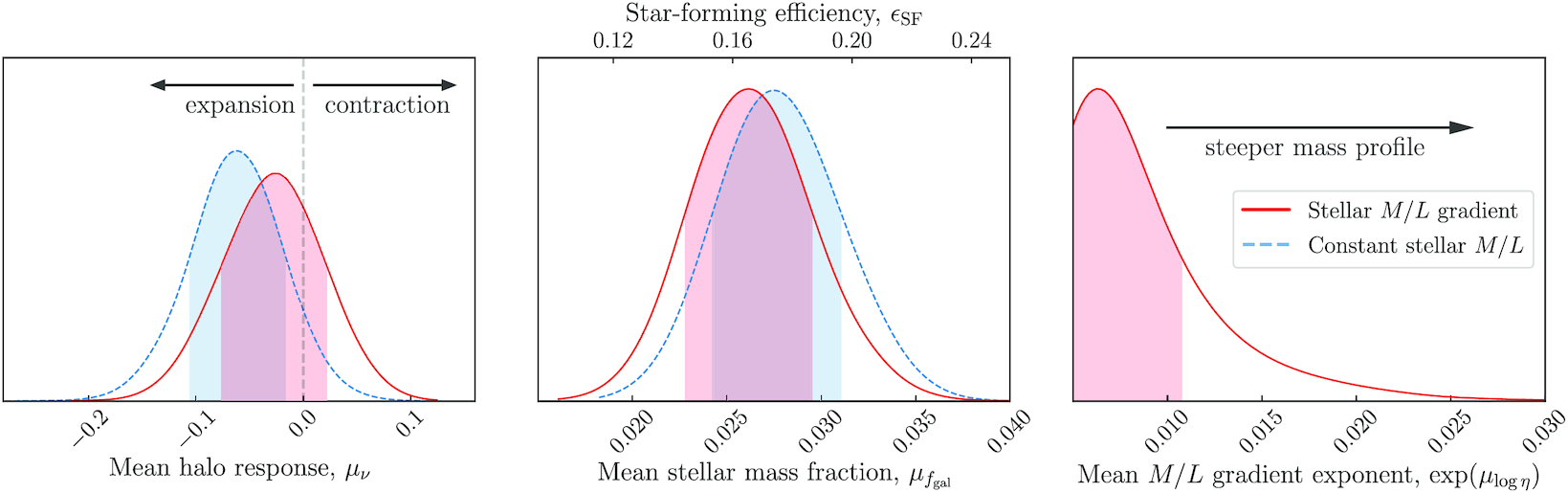}
\centering\includegraphics[width=0.96\textwidth]{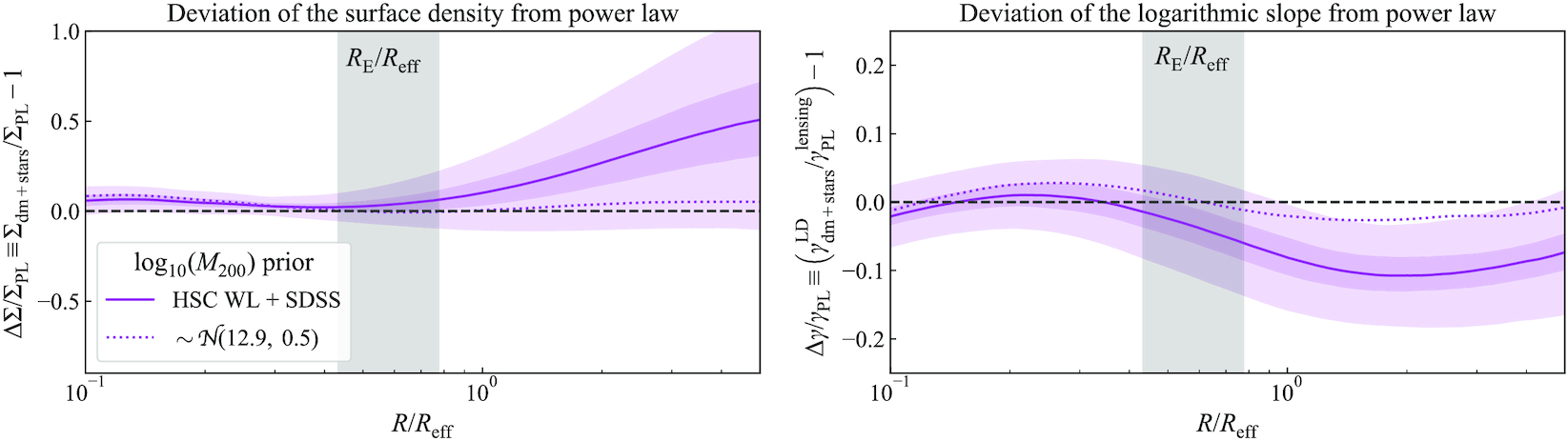}
\caption{Physical interpretation of residual uncertainty allowed by mass-sheet degeneracy on mass density profile.
\cite{Shajib20} modeled a set of non-time-delay lens galaxies with exquisite HST images and unresolved stellar velocity dispersion of the deflector fully accounting for the mass-sheet degeneracy and expressed the results as deviations from standard ``composite'' (top row) and power-law mass profiles (bottom row). The standard composite model comprising a NFW \citep{NFW97} dark matter halo and a stellar component with constant mass-to-light ratio is consistent with the data, although a small amount of contraction/expansion of the halo (top left panel) or a small gradient in mass-to-light ratio (top right panel) cannot be ruled out. Similarly a power-law mass density profile is consistent with the data, although small deviations cannot be ruled out by the data (purple bands in the bottom panels). See \citet{Shajib20} for more description. When available, additional information -- such as spatially resolved stellar kinematics -- reduces the residual freedom and thus tighten the bounds on H$_0$ when applied to time-delay lenses.}
\label{fig:msd-H0-prof}
\end{figure*}

The deflectors' mass distributions are often described using physically motivated models. One frequently used mass model is an elliptical mass distribution with radial power-law profile, where the three-dimensional density $\rho(r)\propto r^{-\rslope}$ with $\rslope$ as the constant power-law slope and $\rslope=2$ as the isothermal profile.  Multiple studies with dynamics, lensing and X-rays have shown galaxies to be well described by a total power-law out to several effective radii \citep[][]{Cappellari15, Gavazzi07}.  Furthermore, the pixelated lens potential reconstruction of the complex interacting lens galaxies in the system B1608+656 showed only small (within $\sim2$\%) deviations from the power-law model, validating the usage of the power-law model for the lens mass distribution, up to a mass-sheet transformation.   A second mass model is a composite of baryonic matter (based on the stellar light distribution assuming a uniform mass-to-light ratio) and theoretically motivated dark matter distribution, typically parametrized as a \citet{NFW97} profile. Multiple studies show that these models also describe well lens galaxies \citep[][and Figure~\ref{fig:msd-H0-prof}]{Shajib:2021}.

Current analysis of time-delay lenses have shown that these two models provide very similar constraints on the time-delay distances within a few percent per lens, and within 1\% from a sample of 6 lenses \citep{Millon20}.  Both mass models have been employed for the primary deflector to quantify uncertainties due to these mass model assumptions \citep[][]{Suyu14, Wong17, Birrer19, Rusu20, Shajib20}, and have been shown to fit the data well.  From a physical point of view one can map residual freedom arising from the mass-sheet degeneracy \citep[][]{Schneider13} as deviations from these baseline models, as shown in Figure~\ref{fig:msd-H0-prof}. 

The profiles of secondary deflectors in the field of view are less crucial as they are not the primary strong lens.  Nonetheless, if they are sufficiently close to the strong lens system, as shown in Figure~\ref{fig:model_eg} panel (a), then these secondary deflectors need to be included in the strong lens model.  In the case where the secondary deflectors are at the same redshift as the primary deflector (e.g. due to natural clustering of galaxies into galaxy groups), all the deflectors can still be modeled via a single lens plane.  The modeling to predict the lensing data $\datalens$ will actually not depend directly on the distances $\dd$ and $\dt$.  In the case where a secondary deflector is at a different redshift than the primary lens, then multi-lens plane mass modeling is required.  Multi-plane lensing requires a cosmological model and redshift measurements of the deflectors in order to trace the light rays that depend on the distance between the different deflector planes \citep[e.g.,][]{McCully17}.  Nonetheless, for the case of HE0435$-$1223 as illustrated in Figure~\ref{fig:model_eg}, one can still define an effective $\dt$ for the system for cosmographic inference despite the secondary deflector having a different redshift from the primary deflector.

The deflector light distribution is perhaps the simplest to model since the \citet{Sersic68} profile  provides a good overall description of galaxy light. Often two or three Sersic components are needed to fit to the primary deflector light distribution \citep{Wong17, Shajib20}.

The AGN host galaxy light distribution can be modeled in several different ways, with the caveat that the model has sufficient flexibility to encapsulate the true underlying light distribution to avoid possible biases in the mass model parameters.  One way is to model the host light distribution on a regular grid of pixels \citep[][]{Warren03, Suyu06} or adaptive pixel grids \citep[][]{Dye05, Vegetti09}. Another way is to use shapelets combined with Sersic profiles \citep[][]{Birrer15}. A recently developed approach is to describe the source through wavelets \citep[][]{Joseph19,Galan21}, although it has not yet been applied to time-delay lens analysis.

With models of the PSF, deflector mass and light distributions, and the AGN host light distribution, we can predict the pixel intensities of the lens system shown in panel (b), which we can then compare to the observations (panel (a)) to compute the lensing likelihood $P(\datalens \vert \parlens, \dt, \dd)$.  The mass model can also predict the time delays between the quasar images in order to compare to the measured time delay (Section \ref{ssec:modern:delays}) and compute the time-delay likelihood $P(\datatd \vert \parlens, \dt, \dd)$.  The primary deflector mass model can be further used to predict the stellar kinematics to compare to spectroscopic measurements $\datakin$, to compute the kinematic likelihood $P(\datakin \vert \parlens, \dt, \dd)$.  With single-aperture averaged velocity dispersion measurements of the primary deflector, which are the typical data obtained from current facilities, often spherical Jeans modeling suffices to fit to the observations \citep[][]{Suy++10, Birrer16}.  Spatially resolved kinematic maps of the deflector would allow us to constrain more flexible mass models and further tighten constraints on the lensing distances for cosmography \citep{Shajib18,Yildirim20,Birrer20,B+T21}. This information was until recently obtainable for only the brighter systems. However, as we discuss in Section~\ref{sec:2020s}, new and upcoming instruments and facilities will provide it for large samples of galaxies.

\subsubsection{Characterizing the line of of sight}
\label{ssec:modern:los}

As described in Section \ref{ssec:MSD}, the presence of external convergence along the line-of-sight affects the inference of $\dt$.  We distinguish between two types of external convergence.  The first type associated with mass structures located sufficiently far (in projection on the sky) from the strong lens system such that their ``flexion shifts'' \citep{McCully17} are small and they can be approximated as mass sheets, with an overall effective external convergence as $\kext$.  In this case, the inferred $\dt$ scales with the factor of ($1-\kext$) via equation (\ref{eq:Dt_mst}) with $\lambda=(1-\kext)$.  The second type of external convergence is associated with nearby massive galaxies located close (typically within $\sim10$'') to the strong lens system.  These galaxies have flexion shifts that are substantial and these galaxies need to be incorporated directly into the strong lens modeling as additional non-linear deflectors, given their non-linear effect on the $\dt$ measurements \citep{McCully17}.  

To incorporate the line-of-sight effects, the first step is to distinguish between these two types of external convergence.  This requires measuring/estimating the redshift of galaxies near the strong lens, estimating the masses of these galaxies through e.g.~stellar population synthesis and relations between stellar mass and halo mass.  It also requires detecting and characterizing any galaxy groups or galaxy clusters near the lens system, as they are thought to have group/cluster scale halos that are more massive than those associated with individual galaxies; these galaxy groups/clusters can have significant flexion shift even if located further away ($\sim$arcmin) in projection \citep{K+Z04}. Wide-field imaging and spectroscopic observations around the strong lens system are therefore necessary \citep[e.g.,][]{Momcheva15, Wilson16, Sluse17, Rusu17, Rusu20}.  

Once we identify the galaxies with negligible flexion shifts (the first type), there are multiple complementary approaches to estimate the external convergence, which include i) using the statistical properties of the environment around the lens system together with large numerical simulations of structure formation, and ii) weak gravitational lensing around the strong lens system. We explain each in turn.

\begin{figure}[t!]
\centering
\includegraphics[width=0.9\textwidth]{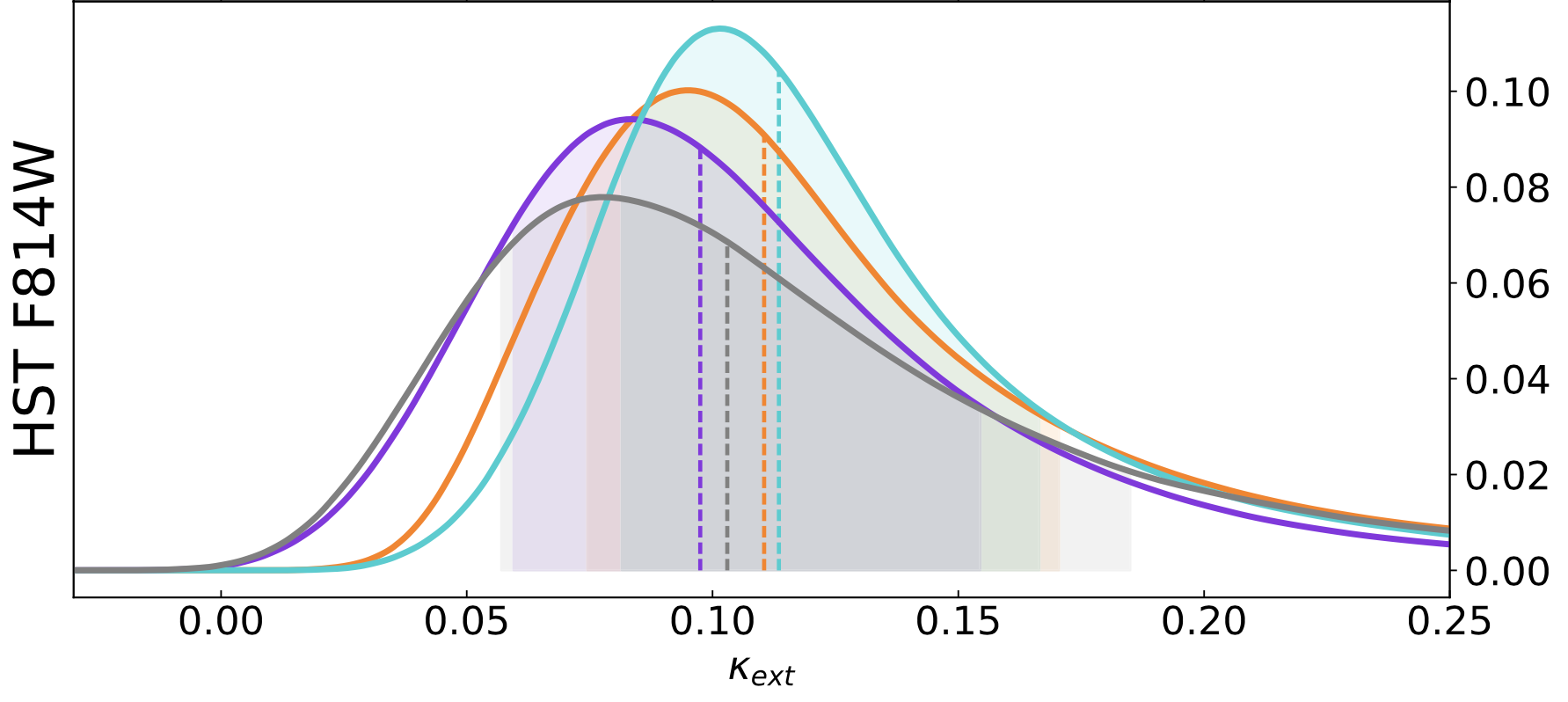}
\caption{External convergence $\kext$ probability density distribution at the gravitational lens system B1608+656 based on HST imaging data in F814W.  Color distributions (orange, purple and blue) are different weak lensing reconstructions using different filters/techniques \citep[][]{Tihhonova20}, whereas the gray distribution is from galaxy counts from \citet{Suy++10}. The $\kext$ distributions from the weak lensing and galaxy counts approaches agree well. Figure panel from \citet{Tihhonova20}. } 
\label{fig:kext_pdf}
\end{figure}

One effective way to characterize the environment around a strong-lens system is through counts of galaxies \citep[e.g.,][]{Fassnacht11} and comparison to those from numerical simulations of structure formation \citep{Hilbert07, Hilbert09}. Lines of sight in the simulations with, e.g., similar relative galaxy number counts as the strong lens system, are then selected and used to reconstruct the distribution of the convergence along the line-of-sight, $\kext$ \citep{Suy++10}.  This method can be extended to more complex summary statistics, such as including weights in the galaxy counts that take into account the mass of the galaxy, the (projected) separation of the galaxy from the strong lens system, and the galaxy redshift, i.e., quantities which affect the strength of lensing and hence $\kext$ \citep{Greene13}.  This approach requires a sufficiently large number of control fields, in order to accurately measure the \textit{relative} galaxy number counts around the strong-lens system; relative galaxy counts are used since they are less sensitive to the inputs of the numerical simulations such as the cosmological parameters than the \textit{absolute} galaxy count. \citep{Rusu17,Rusu20} have optimized this approach in terms of the weights and aperture for obtaining the $\kext$ distribution. Fig.~\ref{fig:kext_pdf} shows an example of the $\kext$ probability distribution reconstructed based on number counts (in gray). 

With high-quality and wide-field imaging, we can also measure the shapes of the galaxies around the strong-lens system and use these shapes to infer $\kext$ through the weak gravitational lensing effect. In short, the shapes of galaxies in different regions of the sky are averaged over, and since (physically unrelated) galaxies' shapes are uncorrelated, any ellipticity in the average shape is due to weak lensing shear by line-of-sight structures.  Measurements of the shear field allows us to reconstruction $\kext$. \cite{Tihhonova18, Tihhonova20} have demonstrated the weak lensing reconstruction for two strong lens systems, HE0435$-$1223 with negligible $\kext$ and  B1608+656 with significant $\kext$. As shown in Fig.~\ref{fig:kext_pdf}, the results of the $\kext$ reconstructed from weak lensing agree well with the $\kext$ obtained from galaxy number counts.

\subsubsection{Putting it all together: blind cosmological measurements}
\label{ssec:modern:cosmo}

By combining the time-delay measurements, the lens mass models, and characterisations of the external convergence along the line of sight, the H$_0$ Lenses in COSMOGRAIL's Wellspring \citep[H0LiCOW;][]{Suyu17} collaboration, together with COSMOGRAIL \citep{Courbin11} and Strong-lensing at High Angular Resolution Programme \citep[SHARP;][]{Chen19}, measured the distances $\dt$ and $\dd$ to 6 lensed quasar systems \citep{Wong20}.  The technique for measuring the distances was developed and demonstrated on the first lens system B1608+656 \citep{Suy++10, Jee19}.  Similar analysis approaches were then applied to the remaining 5 lens systems, done crucially through a blind analysis in order to avoid confirmation bias.  From the posterior measurement of $P(\dt,\dd)$ for each of the lenses (or $P(\dt)$ for two of the lens systems with multiplane strong lensing), constraints on cosmological parameters are inferred through the relations in Section \ref{ssec:TD2COSMO:dist}.

\begin{figure}[t!]
\centering
\includegraphics[width=0.9\textwidth]{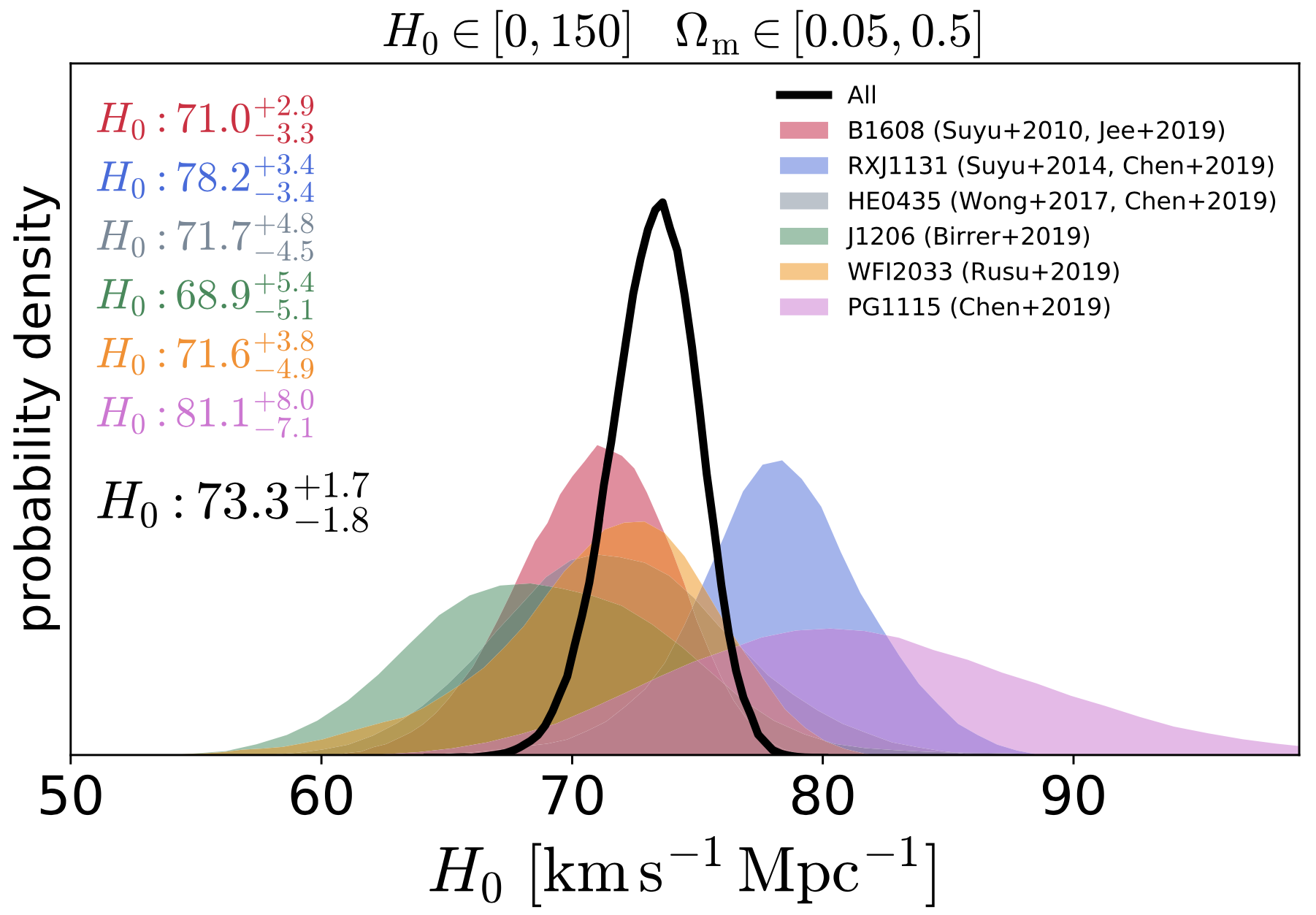}
\includegraphics[width=0.9\textwidth]{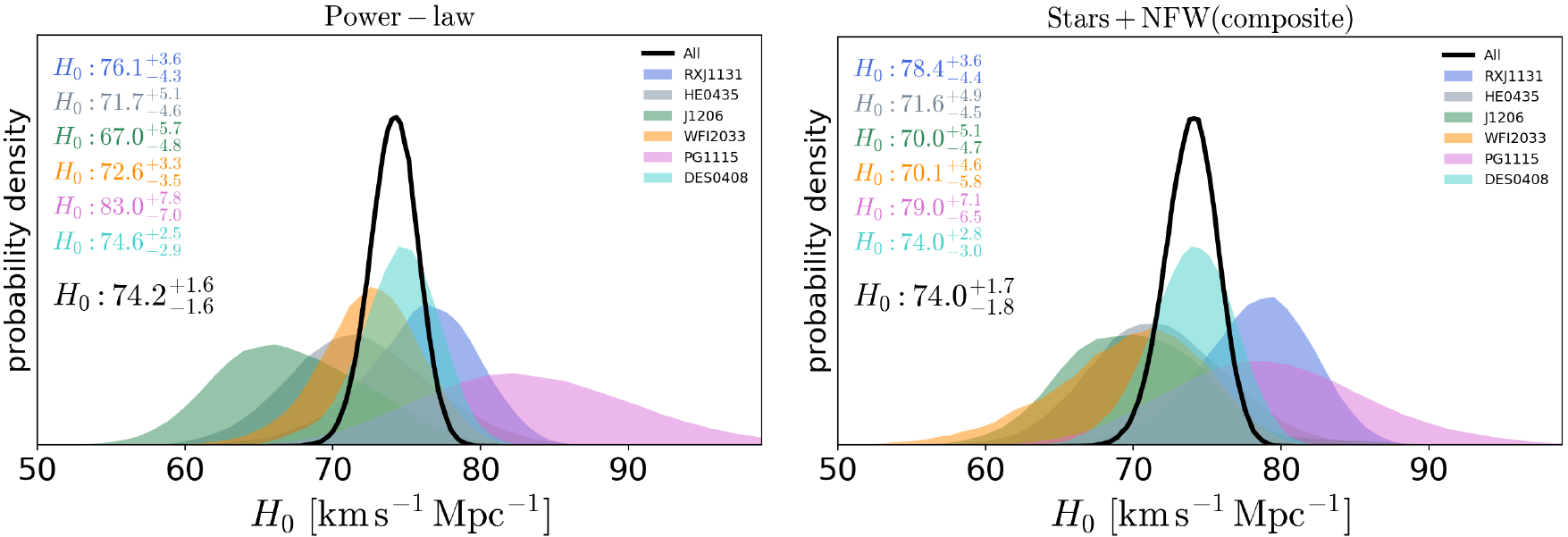}
\caption{Top panel: H$_0$ measurements of 6 lensed quasars from the H0LiCOW program in collaboration with the COSMOGRAIL and SHARP programs, based on well-motivated lens mass models. Figure from \citet{Wong20}.  The joint constraint on H$_0$ in flat $\Lambda$CDM from all the lenses has a relative uncertainty of 2.4\%.  Bottom panels: H$_0$ constraints from the two families of lens mass models, the power-law model (left) and composite model (right) including one additional lens from STRIDES. Figure from \citet{Millon20}.  While the two families of mass models have the flexibility to yield significantly different H$_0$, the two families of models provide H$_0$ measurements that agree within 1\%.}
\label{fig:H0_modern}
\end{figure}

In Fig.~\ref{fig:H0_modern} top panel, we show the posterior probability distribution of H$_0$ from these 6 lensed quasars based on two families of well-motivated models for the primary lens galaxy mass distribution \citep{Wong20, Suy++10, Jee19, Suyu14, Chen19, Wong17, Birrer19b, Rusu20}.  The H$_0$ from each lens are statistically consistent with each other, allowing these lenses to be combined to yield a joint constraint on H$_0$ of $73.3^{+1.7}_{-1.8}\,\kmsMpc$ in flat $\Lambda$CDM, i.e., with 2.4\% relative uncertainty, achieving the goals set out by the H0LiCOW program \citep{Suyu17}.  The STRong lensing Insights into the Dark Energy Survey \citep[STRIDES;][]{Treu18} collaboration has further expanded the lensed quasar sample and measured the distances to a new lensed quasar system in a similar approach, resulting in $H_0=74.2^{+2.7}_{-3.0}\,\kmsMpc$ \citep{Shajib20}. These H$_0$ measurements from strong lensing agree well with that of the local Cepheids distance ladder \citep{Riess21} and are higher than the inference from the cosmic microwave background, making the Hubble tension even more significant.

The Time-Delay COSMOmography \citep[TDCOSMO;][]{Millon20} collaboration, which consists of members of H0LiCOW, COSMOGRAIL, STRIDES and SHARP interested in cosmography with lensed quasars, further investigated systematic effects on H$_0$ measurements, especially given that the mass-sheet degeneracy can manifest itself as degeneracies in the mass profiles \cite[e.g.,][]{Schneider13, Schneider14, Kochanek20, Blum+20}. \cite{Millon20} considered three main sources of uncertainties on the sample of 7 TDCOSMO lenses and found: i) lens stellar kinematic measurements that are systematically offset by 10\% would only change H$_0$ at the 0.7\% level, within the assertive asumptions; ii) no bias on H$_0$ due to incorrect characterization of line-of-sight effects; and iii) the two families of mass models considered have the flexibility to yield significantly different H$_0$ and yet the two families of models provide H$_0$ measurements that agree within 1\%, as illustrated in the bottom panels of Fig.~\ref{fig:H0_modern}.  To conclude, using well-motivated ``assertive'' lens mass models, the current sample of 7 lensed quasars is constraining H$_0$ at the 2\% level.

Alternative approaches to the one of TDCOSMO have also been considered in the literature. For example, pixel based (``free form'') methods have continued to improve in methodology and sample size \citep{Coles2008,Paraficz2010}, although a fundamental issue remains to be addressed. These methods are underconstrained by the data, because typically only quasar positions and time delays are used to constrain the hundreds or thousands of free parameters in the model. Therefore their precision is driven entirely by the implicit or explicit prior on the mass distribution imposed by the regularization schemes. It would be interesting to see this class of models constrained by more information such as extended images or stellar kinematics or, in the case of clusters, large numbers of multiple images \citep{Ghosh2020}. It would also be informative to map the implicit priors to understand what the assumptions corresponds to in terms of physical interpretation.  Finally, incorporating a formal blinding mechanism would help build confidence in these methods.

\begin{figure}[t!]
\centering
\includegraphics[width=0.9\textwidth]{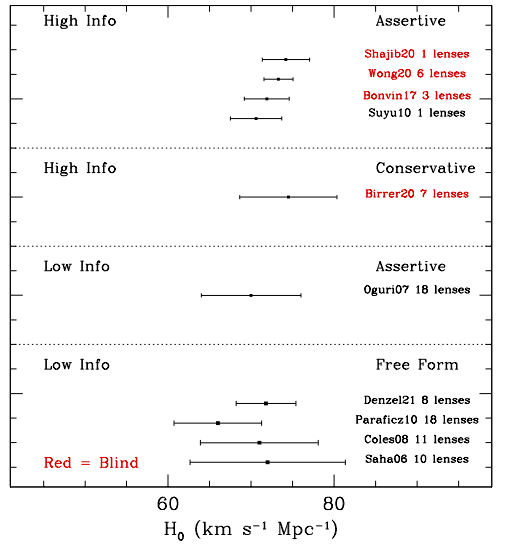}
\caption{Comparison of inferences of H$_0$ based on time delay cosmography, in $\Lambda$CDM cosmology. The measurements are categorized according to i) assumptions on the mass distribution of the main deflector, "assertive" and "conservative" for simply parametrized models as described in Section~\ref{ssec:MSD} or "free form" for pixellated models, as described in Section~\ref{ssec:modern:cosmo}; ii) to the amount of information used per lens, "low info" utilizes quasar positions and time delays, "high info" adds the extended surface brightness distribution of the multiple images of the quasar host galaxy (or "Einstein Ring"), stellar kinematics of the main deflector, and number counts or weak lensing to estimate the line of sight convergence. For each measurement we give the reference and the number of time delay lenses. The measurements shown in red have been blinded to prevent experimenter bias.}
\label{fig:plotH0}
\end{figure}

Another interesting approach is that introduced by \cite{Oguri2007} to exploit samples of lenses with relatively little information per system. By necessity, given the relatively small sample size, these early attempts had to rely on strong informative priors to obtain interesting errors. However, as we will discuss at the end of Section~\ref{ssec:forecasts}, with large enough samples and a proper understanding of the selection function, the hierarchical analysis of many lenses with little information content per system may yield significant precision and accuracy.  Measurements of H$_0$ obtained by the studies discussed in this Section are summarized in Figure~\ref{fig:plotH0}.

\section{Time-delay cosmography in the 2020s}
\label{sec:2020s}

In this Section we review what we consider the main open issues at the moment of this writing (Section~\ref{ssec:open}), describe the considerable opportunities that the coming decade affords (Section~\ref{ssec:opp}), discuss the considerable challenges that will need to be overcome to make the most of these opportunities, from an observational (Section~\ref{ssec:observe}), modeling (Section~\ref{ssec:model}), and
lens-environmental (Section~\ref{ssec:los}) point of view. We conclude by presenting some forecasts.

\subsection{Open issues}
\label{ssec:open}

Even though much has been achieved in the past decade, a number of open issues remain to be addressed for time-delay cosmography to take the next step in precision and accuracy. We discuss the three more important ones in the next subsections.

\subsubsection{``Everything should be made as simple as possible, but not simpler''}

The number one issue is profound, and can be summarized as ``what is a good model?'' Performing a measurement requires a set of assumptions. In a Bayesian framework those include the choice of parametrization as well as the priors on those parameters. In the specific case of time-delay cosmography, as often is the case in astrophysics, each system -- intended as the multiply imaged source and the propagation of photons to the observer -- is inherently complex and cannot be described from first principles. For practical reasons, the problem gets separated into a number of steps: i) the determination of the time delay, ii) the lensing effect of the main deflector, and iii) all the smaller deflections along the line of sight. In turn, each one of these steps needs to be further simplified in order to become tractable. As always, the open issue is how choose the right amount of simplification, or -- as Einstein apparently said in the quote used to title this subsection -- too much simplification leads to underestimated errors and possibly biases. Too little simplification makes the problem intractable, and/or overestimated errors. Ideally, assumptions and simplifications should be a way of expressing empirical knowledge from other sources. For example, a lot is known about quasars and supernovae and their light curves, and that knowledge informs how they are modeled to infer time delays.  
Furthermore, when considering samples of lenses, one should also be mindful of further complexity that arises at the population level. A classic example is understanding the selection function \citep[e.g.][]{Sonnenfeld15,BS21} so that it can be modeled to assess and correct for any potential biases. 

In the case of time-delay cosmography, the number one issue manifests itself most significantly in the modeling of the gravitational potential of the main deflector galaxy. Assumptions of symmetry or of certain functional forms to describe the mass distribution of the main lens affect significantly the inferred precision. If we assume elliptical mass distributions and mass density profiles described by functional forms that are standard and commonly accepted in the literature on massive elliptical galaxies, we infer a 2\% precision on H$_0$ based on current data. If we relax the radial mass profile assumption to make it maximally degenerate with H$_0$, the uncertainty grows to 8\% \citep{Birrer20}. Departures from ellipticity do not seem to be equally important at this stage \citep{VdV22} so one can reasonably conclude that the ``true'' current uncertainty is between 2-8\%. So-called ``free-form'' models based on pixellated mass distribution or potential reach similar conclusions, given reasonable choices of priors and regularization \citep{Denzel21}.
In some sense, this range of uncertainty expresses the difficulty of taking all we know about elliptical galaxies and applying that knowledge to the systems under study. If take the commonly accepted description of the mass profile of early type galaxies, we end up with 2\%. If we know nothing about them, we end up with 8\%. As we will discuss in Section~\ref{ssec:model}, there are two ways to improve what we know about the deflectors of time-delay lenses \citep{Birrer20,B+T21}. The first one is to obtain more data on each one of them. The second is to use knowledge extracted from analogs, assuming that we understand and can model the selection functions. Within the second approach, \citet{Birrer20} demonstrates that the uncertainty can be reduced from 8\% to approximately 5\% if one combines the TDCOSMO sample with the non time-delay lenses from the SLACS sample.

\subsubsection{Perturbations}

The second big open issue is the treatment of the line of sight and nearby perturbers.  The current approach is described in detail in Section~\ref{ssec:modern:los} and yields uncertainties that are currently subdominant with respect to other contributors to the cosmography error budget.

However, there are several subtle effects that need to be studied in more detail if one wants to reach percent precision and accuracy. The first one is how to match exactly the identification of the perturbers in the rendering of the statistical lines of sight.  The comparison light of sights will not reproduce exactly the configuration of nearby perturbers, so one needs to select lines of sight that contribute the same external convergence, except for the closest and most massive perturbers. This is generally achieved by selecting lines of sight that match statistically the observed distribution of galaxies in an annulus around the lens, excising the central part where the individual perturbers are located. It is also required that the simulated lines of sight match the external shear inferred by the strong lens model of the main deflector.  The subtle issues are: i) avoiding double counting in case some of the simulated lines of sight have significant perturbers \citep[see, e.g.,][for a discussion]{Rusu17}; ii) inference of the external shear of the main deflector, introducing covariance between the model of the main deflector, the choice and model of the main perturbers, and the selection of the simulated lines of sight. Another subtle issue is that perturbers are identified as galaxies, and it is very hard to determine whether group-scale halos are present, even with complete spectroscopic catalogs \citep{Sluse17}. A final subtle issue is related to the numerical simulations used for the estimate of the statistical component of the external convergence. In the current scheme one looks at overdensities \citep{Gre++13,Rusu17} and shear \citep{Tihhonova18} to mitigate any dependency of the choice of baryonic physics process in any specific simulation. Furthermore, multiplane ray tracing is performed within the Born approximation. 
As we discuss in the Section~\ref{ssec:los}, in this current decade it will be important to develop a unified scheme to eliminate some of the approximations and carry out diverse cosmological simulations to make sure that the error budget associated with the environment and line of sight remains subdominant as the overall precision and accuracy increase.

\subsubsection{Double Jeopardy: collecting multiple datasets}

The third open issue is logistical. Whereas many experiments in physics and astrophysics are self-contained or based on at most a few instrumental apparatus (e.g. a Cosmic Microwave Background Experiment or a Redshift Survey), time-delay cosmography inherently requires multiple ones. Discovering lenses requires large surveys. Confirming them requires access to a range of telescope for spectroscopy and imaging.
Time-delay determination requires high-cadence and high-precision monitoring, which can be done at relatively low angular resolution ($\sim 1"$) with the assistance of some higher-resolution imaging.
Astrometry and lens modeling require information at $\sim10$ mas scales. Stellar kinematics and redshifts usually require large optical and infrared telescopes, ideally at $\sim100$ mas resolution or better. Characterization of the line of sight requires wide-field imaging and redshifts. Substantial computational power is required to carry out the inferences as well as produce cosmological numerical simulations that can be used for characterization of the selection function and line of sight. How to optimize and muster these resources is an open issue. Each one of these resources is usually allocated independently, and thus double jeopardy is often an issue. Furthermore, allocating committees are usually reluctant to award a piece of the puzzle without guarantees that the others will be secured.  For the determination of time delays, the breakthrough came with guaranteed time on 1-2m class telescopes for multiple years. For the other parts of time-delay cosmography, coordinating and optimizing multiple telescopes remains a big challenge.

\subsection{Opportunities}
\label{ssec:opp}

\subsubsection{Large samples of multiply imaged QSOs and Supernovae}

Large samples of multiply imaged QSOs and supernovae will be transformational for time-delay cosmography. First, the measurements are still in a regime where random uncertainties dominate the error budget. Therefore, increasing the sample of known lenses allows one to shrink the uncertainties. Second, having a large pool of systems to choose from allows one to concentrate the follow-up resources on the ones that yield the biggest ``(cosmological) bang for the observational buck.'' For example, systems with delays estimated to be order 50-100 days are the best suited for time-delay determination with a few percent precision over a single season of monitoring, while it is hard to imagine that such precision can be reached if the time delay is of order a few days. Another example is that systems with a nearby bright star suitable for tip-tilt correction will be easier to target for spatially resolved spectroscopy from the ground using laser guide star adaptive optics. Third, in a landscape where hundreds or more multiply imaged quasars or supernovae are known one can imagine a hybrid approach in which a subset is studied in great detail to inform priors in the modeling of the larger sample.

\begin{figure}[t!]
\centering
\includegraphics[width=0.99\textwidth]{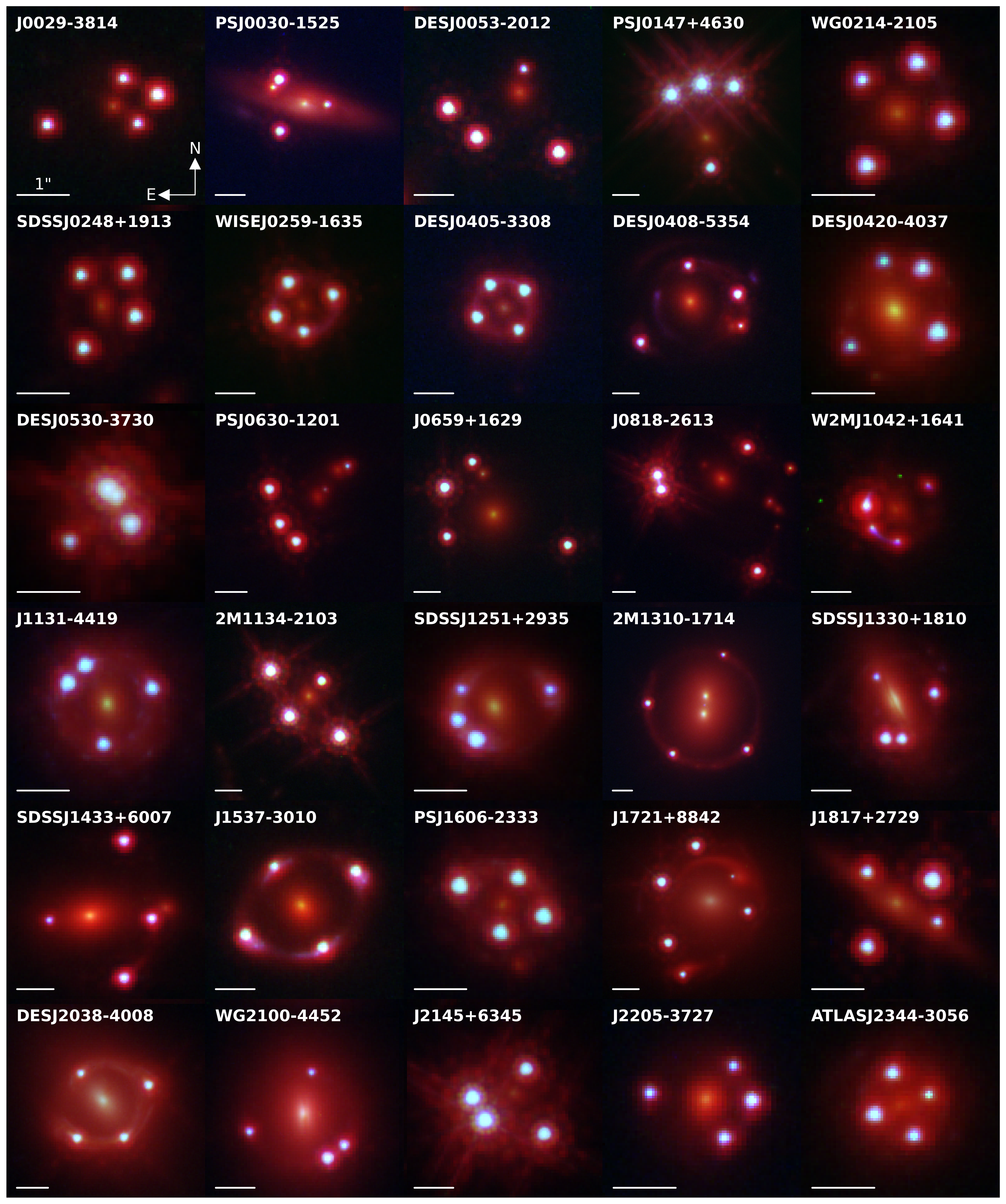} 
\caption{Montage of 30 recently discovered quadruply imaged quasars, imaged with the Hubble Space Telescope and modeled using an automated pipeline. The Figure is taken from \cite{Schmidt22}. The reader is referred to the paper for more details and discovery references.}
\label{fig:30quads}
\end{figure}

\begin{figure}[t!]
\centering
\includegraphics[width=0.99\textwidth]{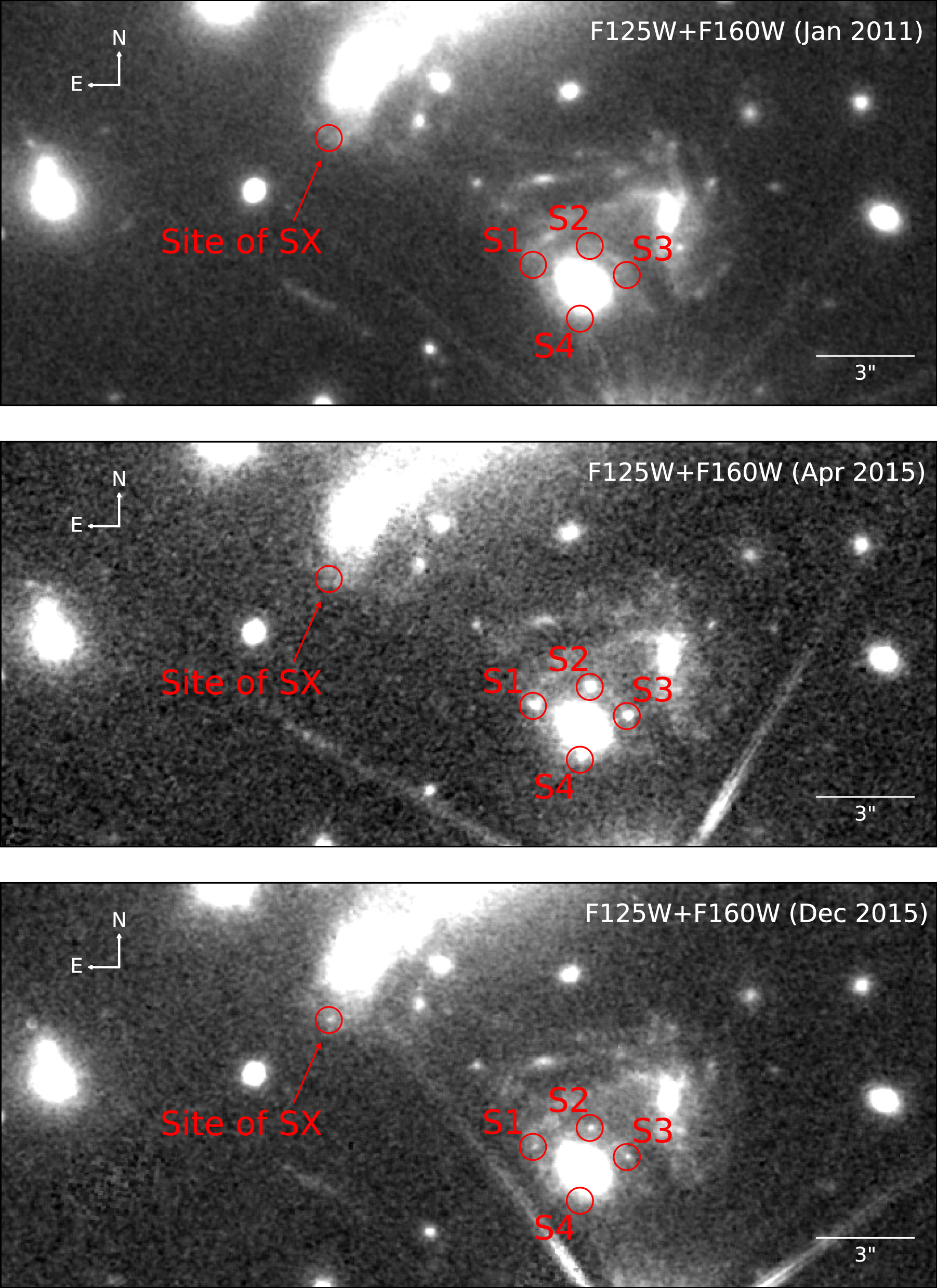} 
\caption{Images of multiply imaged Supernova Refsdal. The time delays between images S1-4 are of order days to weeks, while SX is delayed by approximately 1 year. From \cite{Kelly16}}
\label{fig:Refsdal}
\end{figure}

As reviewed in Section~\ref{ssec:modern:sample}, multiply imaged quasars and supernovae are rare in the sky. Only a few hundreds of the former \citep{Lemon2020,Lemon2022} and a handful of the latter \citep{Kelly15, Goobar17} have been found. Examples of discoveries since TM16 are shown in Figures~\ref{fig:30quads} and Figure~\ref{fig:Refsdal}. 

A novelty of the 2020s will be the discovery and exploitation of multiply imaged supernovae for time-delay cosmography. At the moment of this writing, the first determination of a high-precision time delay for a multiply imaged supernova and resulting determination of the Hubble constant has been submitted for publication, and we expect many more to follow \citep[e.g.,][]{Goldstein19, Wojtak19, Huber19, Grillo20, Suyu20}. Lensed supernovae present some advantages and disadvantages with respect to the lensed quasars from the point of view of time-delay cosmology. The main disadvantage is that by being transient they require cadenced surveys or monitoring campaigns to be discovered and cannot be found in single-epoch data. The main advantages are: i) supernova light curves are not stochastic like those of quasars and thus one can use templates or models to infer the time delay requiring less demanding models \citep[e.g.,][]{Pierel19, Huber22}; ii) supernova spectra evolve over time and can be used as a new way to measure the time delays \citep{Johansson21, Bayer21}; iii) some supernovae (e.g. Ia) are standardizable candles, which gives an independent handle on the absolute magnification and therefore a way to mitigate the mass-sheet degeneracy, provided that microlensing/millilensing magnifications of the supernova can be quantified accurately \citep{OK03,Foxley-Marrable18, Yahalomi17,BDS22}; iv) after the transient has faded, its host galaxy and the foreground detector can be studied more cleanly than in lensed quasars where the light from the quasar itself is often a dominant source of noise \citep{DingLiao2021}. 

In the 2020s more powerful facilities and surveys will enable order of magnitude increases in the sample size of both multiply imaged quasars and supernovae \citep{O+M10,Yue22}.  The Legacy Survey of Space and Time (LSST) of the Vera C.~Rubin Observatory, scheduled to start operations
in 2024, will image approximately 18,000 square degrees of sky with depth and cadence vastly superior to that achieved by current state-of-the-art surveys. LSST will not only deliver large numbers of lenses by virtue of the large survey area, but it will also enable the discovery of lenses via variability \citep{Pindor2005,KMMS06, Chao20,Bag22,KodiRamanah22} and the determination of some time delays from the survey data itself \citep{Liao15}. 

Euclid and then the Nancy Grace Roman Space Telescope -- scheduled to be launched in the next few years -- will survey thousands of square degrees at angular resolution $0.2"-0.1"$, more than sufficient to resolve multiple images for typical Einstein radii of order arcsecond. The image quality will be sufficient for direct confirmation of the multiply imaged sources, bypassing the higher-resolution imaging follow-up that is often needed to confirm the candidates identified in seeing limited data. 

Two other techniques could provide additional lenses. On the one hand, spectroscopic surveys with the Dark Energy Spectroscopic Instrument \citep{DESI}, 4MOST, MOONS, and the Prime Focus Spectrograph \citep{PFS} will measure millions of spectra that can be searched for double spectra, identifying strong lensing systems on much larger scales than possible in the past \citep{Huchra1985,Bolton06,Bolton08}. One of the advantages of spectroscopically identified strong lensing systems is that the redshift information is available from the survey itself.

On the other hand, even unresolved time domain data can be used to identify lensing events. Unresolved variable multiply imaged sources will appear as light curves with multiple autocorrelation peaks \citep{Shu21, Bag22} corresponding to the time delays. This technique is in principle applicable with relatively small telescopes of large etendue such as ZTF \citep{ZTF} or LAST \citep{LAST}.

\subsubsection{External datasets}

At any given time, the vast majority of galaxies does not host an active nucleus or a supernova. Therefore the large surveys planned for the 2020s will deliver much larger numbers of so-called galaxy-galaxy lenses than quasar-galaxy or supernova-galaxy lenses. Galaxy-galaxy lenses do not contain the absolute distance information that comes from the time delays, but they can be used to inform how we model the quasar-galaxy or supernova-galaxy systems \citep{Birrer20,B+T21,Sonnenfeld2021}.
The sheer number of such galaxy-galaxy lenses to be found in the surveys under way in the 2020s all but guarantees that crucial data (e.g. high-resolution imaging, redshifts of the source, and velocity dispersion and redshift of the deflector) will be available from survey data themselves. 

\subsubsection{Integral field spectrographs}

The launch of the James Webb Space Telescope and the constant upgrade of adaptive optics systems on 8-10m (and Extremely Large Telescopes, 25-39m, in the near future) class ground-based telescopes imply that our ability to do resolved spectroscopy at the angular resolution relevant for strong lensing systems ($0.1-0.2"$) will be unprecedented in the 2020s. Spatially resolved kinematics has been recognized for some time as a fundamental ingredient to break the mass-sheet degeneracy and the mass-anisotropy degeneracy \citep{Shajib18,Yildirim20}, but it was extremely challenging with 2010 technology. There is reason to believe the 2020s will be revolutionary in this respect.

\subsection{Observational challenges}
\label{ssec:observe}

In order to make the most of the large sample of lenses to be discovered in the 2020s, two observational challenges will need to be solved. 

The first one is the bottleneck of follow-up. Time-delay cosmography requires multiple pieces of information, and the surveys will not provide all of them; follow-up campaigns will be required. We anticipate that high-cadence targeted monitoring will be needed in most cases to determine time delays especially for lensed supernovae \citep{Huber19} (either by supplementing survey light curves or from scratch for systems discovered in single-epoch data). Most of the new lenses to be discovered in the 2020s will be fainter than the ones monitored prior to 2020, owing to a combination of the rising number counts towards fainter magnitudes and the fact that the brighter systems have likely already been discovered across most of the sky \citep{Treu18}. Thus, even monitoring just 50-100 lenses to obtain exquisite time delays will require dedicated 3-4m class telescopes, as opposed to the 1-2m class telescopes that were used with great effect prior to 2020. Furthermore, most survey data will be of insufficient depth and resolution to perform cosmography-grade lens models \citep{Meng15}. Thus, follow-up with the Hubble Space Telescope (as long as it is operational), JWST, adaptive optics from the ground, or radio interferometers when applicable, will in most cases be needed. Fortunately, single-epoch imaging will be sufficient for this task (or at most twice for transients, during and after), meaning that it is feasible with existing and planned facilities.
Finally, as mentioned above, spatially resolved kinematics for large samples of lenses will be transformative. What is needed is continued support for the construction and deployment of the facilities \citep{LIGER} and the allocation of sufficient telescope time.

The second challenge will be understanding and characterizing the selection function of each sample with sufficient degree of accuracy. It is well established that the most common deflectors, massive elliptical galaxies, are a very uniform population that can generally be characterized by just two parameters, as evidenced by the tightness of the fundamental plane \citep{D+D87,Dre++87a} and mass plane \citep{Auger10}. This uniformity is a good starting point for constructing matched samples of time-delay and non-time-delay lenses \cite[e.g. by matching velocity dispersion and redshift,][]{Tre++09} and should mitigate biases related to lensing selection \citep{Collett16,Lutharu}. However, detailed and specific calculations of the selection function for each sample  will be needed in order to meet the 1\% precision and accuracy goal. Those calculations will need to consider every step of the process contributing to time delay cosmography, from lens system identification and confirmation, to time delay measurement, and lens and environment properties, and quantify any potential bias arising from these steps.

\subsection{Deflector-modeling challenges}
\label{ssec:model}

In the upcoming decade,
we expect to have approximately $\sim10^5$ lensed galaxies, $\sim10^3$ lensed quasars, and $\sim 10^2$ lensed supernovae \citep{Collett15, Oguri10, Wojtak19} from LSST and/or Euclid.  We would have: i) a large sample with only LSST data, ii) a medium sample with LSST and Euclid images, and iii) a small sample with LSST, Euclid and ancillary follow-ups necessary for cosmography including HST/JWST images and spatially resolved kinematics.  The challenge is to make use of all the lenses for cosmography, by for example learning about the mass structure of galaxies from the large sample and using it as prior for the small sample.  This requires both enhancement in the speed of the modeling, and also new development in the modeling techniques especially with respect to joint lensing and dynamical models of galaxies.

The conventional modeling approach described in Section~\ref{ssec:modern:modeling} takes typically at least weeks per lens, and will therefore be far too time-consuming for the large and medium samples of $\gtrsim 10^3$ lens systems.  \cite{Hezaveh17, PerreaultLevasseur17}  pioneered the use of deep learning to infer lens mass model parameters quickly at $<1$\,second per lens after training the neural network for lensed galaxy systems with HST images.  Various studies are investigating this approach for ground-based images and also future Euclid-like and LSST-like images \citep[e.g.,][]{Pearson19, Schuldt21, Pearson21}, and for its compatibility with the hierarchical inference framework introduced above \citep[][]{Wagner-Carena2021,Park2021}.  In addition, the use of Graphics Processing Units (GPU) can save substantial computational time \citep{Gu22}. The challenges are to validate these new modeling methods on real lenses since these methods have so far been demonstrated only on mock lenses.  First steps in this direction are being taken (Schuldt et al., in prep., Erickson et al., in prep.).  In contrast to the large/medium samples, the small sample of $\sim50-100$ lensed quasars with necessary follow-up is small enough that the conventional modeling approach is feasible in terms of computational time, but reduction in the human investigator time per lens system is still needed to analyze the small sample.  This could be done by automating the modeling procedure \citep{Shajib19,Schmidt22,Ertl22} or further developing the use of GPUs or deep learning as described previously.  

In addition to the needed enhancement in the modeling speed, new frameworks for lensing and dynamical mass models are required, especially with the upcoming breakthrough in acquiring spatially resolved kinematic maps of the lenses for the first time.  In the past decades, only single-aperture-averaged velocity dispersions of the foreground lens galaxy have been obtainable for lensed quasars; spherical Jeans modeling, while adequate and often used for modeling such data, will no longer be sufficient for the JWST era where we expect high-SNR spatially resolved kinematics. \cite{Barnabe11, Barnabe12, Yildirim20} have developed self-consistent lensing and dynamical modeling, and further work in this direction is warranted to fully exploit the next generation of data.

To uncover any systematic biases in the mass modeling, multiple independent models invoking a wide range of plausible lens mass distributions are needed.  The recent blind modeling of the same lens system WGD\,2038$-$4008 by \citet{Shajib22} with two independent modeling softwares show good agreement in the predicted time delays within 1.2$\sigma$.  However, the radial mass profile slope between the two models differ significantly and is mostly attributed to the unknown point spread function (PSF), which is reconstructed in the modeling procedure \citep[e.g.,][]{Chen16, Wong17, Birrer19b}.  With the next generation of facilities especially with ground-based AO systems, new developments in characterizing the PSF accurately are worth pursuing to enable even more accurate lens mass reconstruction for cosmography. 

\subsection{Line-of-sight-modeling challenges}
\label{ssec:los}

The two approaches to estimate the LOS external convergence described in Section \ref{ssec:modern:los}, galaxy number counts and weak lensing, rely on simulations of the large-scale structure.  The ray tracing through the Millennium Simulation \citep{Hilbert07, Hilbert09}, a dark-matter-only simulation with semi-analytic galaxies, has been the main simulation used for this purpose.  While the effects of baryons are expected to be small for the large-scale structure associated with the LOS external convergence, these could become important for getting H$_0$ accurate to the percent level.  Quantifying the effects of baryons using new hydro-simulations of dark matter and baryons on large scales is worth pursuing.

A potential way to further improve the external convergence determination is to reconstruct the entire light cone of structure around the strong lens, by assigning mass to the galaxies that are observed.  
First theoretical frameworks have been put forth \citep[e.g.,][]{Collett13, McCully17, Fleury21}.  
Subtleties remain in decoupling the cosmological dependence of the external convergence in order to measure the $\dt$ and $\dd$ distances without dependence on cosmological model assumptions.  
Further developments of these new directions are warranted. 
In general, the most challenging aspect of this approach is the high dimensionality of the model: in principle, the mass, redshift, luminosity and other properties of the galaxies in the light cone should all be free parameters needed to fit the observed noisy catalog measurements and predict, with meaningful uncertainty, the small-scale weak lensing effects at the lens position. 
Graph neural networks could be an efficient way to use all the available catalog data and produce (approximate) predictions in finite time (Park et al, in preparation). 
It remains to be seen whether using more information in this way yields a method that out-performs the number counts summary statistic approach in use today.

\subsection{Other challenges}
\label{ssec:other}

Another challenge is given by the social aspect of science, and it is common to other areas of astrophysics and cosmology as they move from idea, to proof of concept, to large experiment scale. At the beginning, a single person \citep{Ref64}, or a small group of scientists \citep{Suy++10}, are enough to have the idea and carry out the proof of concept. However, in order to meet the target 1-2\% precision and accuracy on H$_0$ that are needed to contribute to the debate on the ``Hubble tension,'' large-scale efforts are needed, which inevitably lead to larger teams \citep{Treu18,Wong20}, partly to deal with the diversity of telescopes discussed above and partly just to carry out the analysis. Inevitably, scaling up requires more time to be devoted to project management, communication, collaboration infrastructure, and quality control. The latter is particularly challenging because no single individual can be an expert in all the myriad of tasks and therefore mechanisms needed to be put in place to perform quality control over a distributed effort, in addition to the need to preserve the blindness of the measurements. There is also the need to ensure that work in all aspects of the measurements, not just on deriving the final numbers, is properly rewarded, especially that of junior collaborators, and of those who spend time in building the ``social infrastructure.'' These issues are not unique to time-delay cosmography and have been addressed in other contexts but are certainly of paramount importance if one wants to make the most of the opportunities in the 2020s.

\subsection{Forecasts}
\label{ssec:forecasts}

We conclude our review with a set of forecasts for what we can expect time-delay cosmography to achieve before this decade is out. We consider three scenarios: i) optimistic; ii) reasonable; iii) pessimistic. 

The optimistic scenario postulates that all the new observational facilities listed above are deployed approximately according to schedule, follow-up resources are available, and that methodological improvements are sufficient for the analysis to keep up with the pace of the data acquisition.
The reasonable scenario is a middle ground in which some of the facilities will be on time and some will be delayed. Follow-up data will be acquired, but occasionally they will be delayed, and at least some of the methodological improvements will bear fruit. In the pessimistic scenario we will assume no new facilities nor methodological improvements, i.e. applying only existing and demonstrated technology.

For each of the three scenarios we will consider two sets of assumptions about the radial profile, which should bracket the true expected uncertainty, as discussed before. The most restrictive assumptions (``assertive'' in the language of Section~\ref{ssec:MSD}) correspond to knowing the form of the standard mass density profile up to 2-3 free parameters, informed by observations of galaxies or by cosmological numerical simulations. It is the generalization of the approach carried out by the H0LiCOW/SHARP/STRIDES teams up until the papers by \cite{Wong20,Shajib20,Millon20}. The most generic assumptions (``conservative'' in the language of Section~\ref{ssec:MSD}) correspond to a parametrization of the radial mass density profile that is maximally degenerate with H$_0$ through the mass-sheet transform \citep{Birrer20}. We will also assume that systematics (except the mass density profile assumption) are below our target 1\% level. For the restrictive assumption case, we assume that each lens will deliver a time delay distance measurement as precise as the average precision over the sample of 7 lensed quasars studied by \citet{Wong20,Shajib20,Millon20}.
For the generic assumptions we will combine information from time-delay lenses and external datasets following \citet{B+T21}. 

In the optimistic and reasonable scenario there will be an abundance of lensed quasars and supernovae to choose from, and progress will be limited by the pace of follow-up and modeling. An activity driving the schedule will be time-delay measurements. For this, we shall assume that some dedicated 3-4m class telescope or multiple 2m class telescopes are available in both scenarios. In the optimistic scenario the Vera C.\ Rubin Observatory's LSST will start contributing time delays in large numbers towards the end of the decade (likely with supplemental monitoring to increase cadence), accelerating the progress further. For both scenarios, we will assume that stellar kinematics can be derived for all the relevant targets using either JWST (20\% of the targets) or AO-assisted ground based telescopes (80\% of the targets). In the optimistic scenario a further boost will arrive from Extremely Large Telescopes at the end of the decade to deal with LSST's increased pace of time-delay measurements. For both scenarios we will assume that external datasets as defined by \citet{B+T21} will be available and that analysis team will keep up with the data rate (5/yr for reasonable, gradually increasing to 20/yr in 2032 for optmistic).  

In the pessimistic case, discovery pace will slow down progress further. Current samples already include 40 quads, and with existing facilities the numbers are bound to increase. However, without the ability to optimize the choice, progress will be slower partly for logistical reasons (e.g. monitoring efficiency for a given dedicated telescope) and for scientific reasons (e.g. highly symmetric configurations have very short delays that cannot be measured to sufficient precision with current technology).  We will thus assume a constant rate of 3 new time-delay systems completed per year, which is a sustainable rate at current level of effort and technology. As in the other cases we assume that spatially resolved stellar kinematics is available from JWST (20\%) or ground-based AO-assisted spectrographs (80\%).  It is important to notice that this a conservative forecast in the sense that it focuses on quads. Doubles vastly outnumber quads \citep[e.g.,][]{Lemon2022}. Therefore if sufficient resources can be found to expand monitoring, modeling, and stellar kinematic measurements to doubles, they would provide a significant boost to sample size over a short time scale.

\begin{figure*}[!t]
\centering\includegraphics[width=0.96\textwidth]{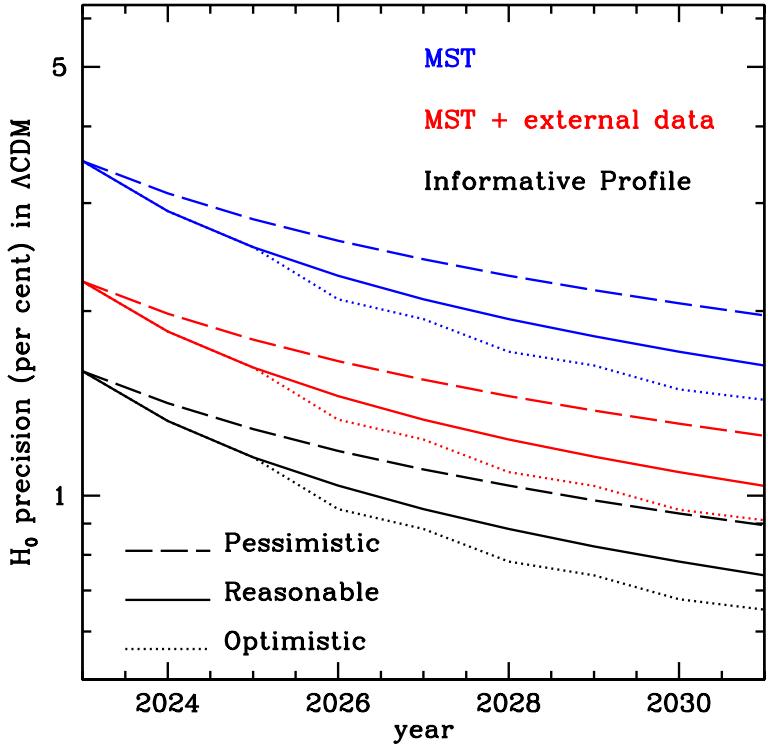}
\caption{Forecasted precision on H$_0$ under a range of assumptions in terms of discovery/analysis rate (pessimistic to optmistic) and in terms of the mass density profile of the main deflector (informative or ``assertive'' vs ``conservative'' or maximally degenerate with H$_0$ through the mass-sheet transformation (MST), with and without external data). We assume spatially resolved kinematic data is available for all time-delay lenses (80\% from ground-based AO and 20\% from JWST) but not for the external datasets. The definition of the terms is given in Section~\ref{ssec:forecasts}. The forecast does not include a possible systematic floor that may arise from inaccuracies in the correction of biases stemming from the selection function, or from currently "unknown unknows".}
\label{fig:roadmap}
\end{figure*}

The resulting forecasts are shown in Figure~\ref{fig:roadmap}, which is by design similar in spirit to the one showed by \cite{Treu16b}. In all cases we start from a sample of 11 in 2023, which is the currently published seven plus four that are in advanced stage of modeling at this time. The ``Informative Profile'' track is the one directly comparable with the forecast of 2016. It is sobering to notice that progress so far has been slower than expected back then. This is due to primarily two factors. First, delays in facilities and surveys has reduced the number of accessible targets and follow-up. Second, the deliberate focus on systematics during 2019-2022 have slowed down the analysis of new systems. The lessons have been incorporated in the forecasts presented here which are significantly more conservative in terms of pace and also include the conservative assumptions accounting for the MSD. Based on the updated knowledge it seems that 1\% is within reach, although it will require exploitation of external samples in the conservative approach. Regarding external samples, our forecast has been following \cite{B+T21}, requiring full information for all the non-time-delay lenses, but it has been proposed that much larger samples with less information per system could also improve the yield \citep{Son21a,Son21b,Son22a,Son22b}. This avenue would further boost the expected precision.

\begin{figure*}[!t]
\centering\includegraphics[width=0.96\textwidth]{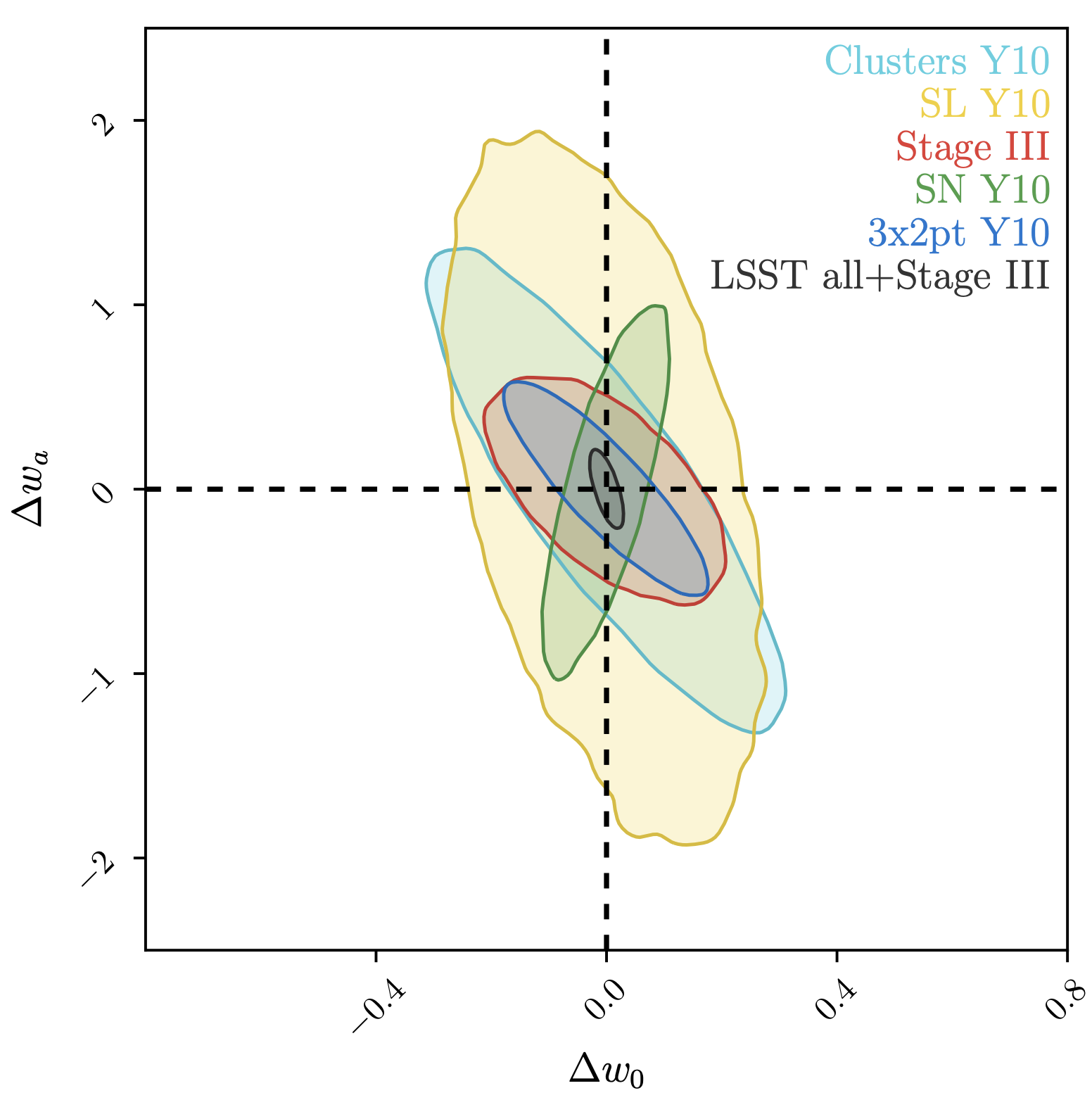}
\caption{Strong lensing as a competitive LSST Dark Energy probe. This excerpt from Figure G2 of the LSST Dark Energy Science Collaboration's Science Requirements Document \citep{LSSTDESC18} shows the potential of time-delay cosmography (yellow) to contribute to the joint constraint of Dark Energy equation of state parameters. The assumptions in this forecast are described in the text. While the contours from the different ``Y10'' probes are aligned by construction, one can see that any tension that emerges between any two of the probes should be able to be well-tested with the other two.} 
\label{fig:desc-srd}
\end{figure*}

The above forecast, and indeed most of the discussion in this review, has all been focused on H$_0$, as the cosmological parameter best constrained by time-delay lenses (and of highest current interest). However, we should keep in mind that time-delay distances can be used in general as a probe of Dark Energy that is highly complementary with other ones \citep{Lin11}. 
We refer the reader to the LSST Dark Energy Science Collaboration's (DESC's) Science Requirements Document (SRD) \cite{LSSTDESC18} for their 2019 forecast of the ability of the LSST strong lens sample to contribute to the overall measurement of the Dark Energy equation of state parameters;  we reproduce an excerpt from one of the key figures from that work in Figure~\ref{fig:desc-srd}. The DESC SRD forecast predated the last three years' analysis of modeling systematics, and so is both on the optimistic and assertive side (in the language of the preceding discussion). It assumes, relative to Figure~\ref{fig:roadmap}, a larger number of systems (400) that have both measured time delays and high-resolution imaging data (citing the Euclid and Roman surveys as possible sources for the latter, and following the results of the Time Delay Challenge \cite{Liao15} for the former) will have been assembled by year 10 of the LSST survey (2035); on the other hand, it only anticipates 7\% distance precision per lens, and does not include any lensed supernovae. This forecast (and any subsequent update to it) provides a valuable rallying point for those looking to ensure that time-delay cosmography contributes to the measurement of Dark Energy in the 2020's and 2030's in a meaningful way: it demands a significant step up in the rate at which well-measured time-delay lens systems are added to the sample, and motivates the investigation of larger samples that have survey data alone. It should be noted that additional constraining power on dark energy from strong lenses may come from double source plane lenses \citep{Gavazzi08,Collett12} without time delays, which are not, however, the subject of this review.
 
\section{Summary}
\label{sec:summary}

Strong lensing gravitational time delays have been demonstrated to be a powerful tool for cosmography. By measuring absolute distances they provide a way to determine the Hubble constant, independent of all other methods, and can help settle the debate over the so-called Hubble tension.

The 2020s have the potential to be transformative for time-delay cosmography. Technological and methodological progress can lead to substantial improvements reaching $\sim$1\%  precision and accuracy on the Hubble constant and parallel improvements in the determination of other cosmological parameters.

In this article, we briefly review the foundations of the method and recent progress in its application before turning to a critical discussion of prospects for the 2020s. We identify the following open issues, opportunities, and challenges for the coming decade.

The three open issues to be solved are as follows. First, ``what's the right degree of flexibility in lens models?" Too much flexibility and one is throwing away information, too little and one is underestimating the uncertainties. The current approach consists of bracketing the real answer with ``assertive" and ``conservative" assumptions, but it is clear that a way to make clear progress is to inject more information, either from non-lensing information such as spatially resolved kinematics of the deflectors, or from external datasets (galaxies that are not time-delay lenses or not lenses), or from numerical cosmological simulations if they can reach the desired level of fidelity. Second, ``how to deal with perturbers?", intended as every mass component along the line of sight excluding the main deflector. The current approach consists of separating the  
weak perturbers and accounting for them statistically, while modeling explicitly the more significant ones. At the moment this treatment is sufficient, but the issue needs attention in the next few years if one wants to reach the accuracy and precision goal. The third issue is logistical. Time-delay cosmography requires multiple pieces of information coming from a variety of telescopes. This makes acquiring all the necessary data difficult and stochastic in a world where time is allocated by telescope. Solving this problem will probably require a combination of dedicated facilities (e.g. for time delays) and an expansion of mechanisms for multi-telescope allocations, which is so far very rare in astronomy (e.g. Hubble and Chandra).

The opportunities offered by the 2020s are immense. Samples of lensed quasars and supernovae are going to explode in size, allowing observers to choose the best targets for time-delay cosmography maximizing the impact per object. The same surveys that will discover new lensed quasars and supernovae are also going to discover many more non-time-delay lenses, which could be a decisive factor in boosting accuracy and precision. Likewise the surveys are going to provide a substantial amount of the data needed for the analysis -- although crucially not all, in general. Finally, the improvement of optical-infrared integral field spectrographs on the ground and in space should really be transformative, finally making spatially resolved stellar kinematics possible for large samples of lenses.

In order to take advantage of the opportunities, the community will have to overcome a number of challenges. Observationally, the main challenge will be to gather all the necessary follow-up. We see time delays, high-resolution imaging, and spatially resolved kinematics as likely bottlenecks for follow-up. For time delays, having dedicated 2-4m class telescopes to supplement survey cadence seems a high priority. JWST and the planned space missions, plus the continuing development of adaptive optics on large ground-based telescopes (8-39m), are the priority for high-angular-resolution imaging and spectroscopy. In order to take full advantage of this wealth of data, faster accurate lens modeling is a high priority, alongside more detailed studies of the line-of-sight perturbers. Public challenges to validate the methods \cite[building on][]{Liao15,Ding21} will continue to be an important tool to test for accuracy.  

We conclude by attempting some forecasts for time-delay cosmography. We consider scenarios ranging from pessimistic to optimistic. We conclude that, even under conservative assumptions regarding our knowledge of the mass density profile of massive ellipticals, the $\sim$1\% precision and accuracy goal on H$_0$ is within reach, provided that we exploit the knowledge provided by external datasets and we are able to collect all the necessary data for a sample of $\sim$ 40 quads (QSO or SN). The same sample will serve as a powerful probe of dark energy and other cosmological parameters in combination with other more traditional probes. The rate of progress may be accelerated by the study of larger samples of doubles, if sufficient resources can be mustered to monitor, model them and obtain the necessary ancillary data.

\bmhead{Acknowledgments}

The authors thank their TDCOSMO colleagues for many insightful conversations over the years. The authors thank Simon Birrer, Xuheng Ding, Liz Buckley-Geer, Shawn Knabel, Anowar Shajib, Paul Schechter, Liliya Williams for helpful comments on a first draft of the manuscript. The referees are grateful to the referees for their constructive reports that help improved the manuscript. TT acknowledges support by the National Science Foundation through grants 1906976 and 1836016, by the National Aeronautics and Space Administration through grants HST-GO-15652 and JWST-GO-1794, and by the Gordon and Betty Moore Foundation through grant 8548. SHS thanks the Max Planck Society for support through the Max Planck Research Group.  This project has received funding from the European Research Council (ERC) under the European Union's Horizon 2020 research and innovation programme (grant agreement No.~771776). This research is supported in part by the Excellence Cluster ORIGINS which is funded by the Deutsche Forschungsgemeinschaft (DFG, German Research Foundation) under Germany's Excellence Strategy -- EXC-2094 -- 390783311.

\end{document}